\documentclass[a4paper,fleqn,usenatbib]{mnras}

\usepackage{newtxtext,newtxmath}

\usepackage[T1]{fontenc}
\usepackage{ae,aecompl}

\usepackage{graphicx}	
\usepackage{amsmath}	
\usepackage[export]{adjustbox}

\title[MeerKAT discovery of the Vela X-1 bow shock]{MeerKAT discovery of radio emission from the Vela X-1 bow shock}

\author[Van den Eijnden et al.]
{J. van den Eijnden,$^{1}$\thanks{E-mail: jakob.vandeneijnden@st-hildas.ox.ac.uk} 
I. Heywood,$^{1,2,3}$ 
R. Fender,$^{1,4}$
S. Mohamed,$^{3,4,5}$ 
G. R. Sivakoff,$^{6}$ 
\newauthor 
P. Saikia,$^{7}$ 
T. D. Russell,$^{8,9}$ 
S. Motta,$^{1,10}$ 
J. C. A. Miller-Jones,$^{11}$ 
P. A. Woudt$^{4}$
\\
$^{1}$Astrophysics, Department of Physics, University of Oxford, Denys Wilkinson Building, Keble Road, Oxford OX1 3RH, UK\\
$^{2}$Department of Physics and Electronics, Rhodes University, PO Box 94, Makhanda, 6140, South Africa\\
$^{3}$South African Radio Astronomy Observatory, 2 Fir Street, Observatory, 7925, South Africa\\
$^{4}$Department of Astronomy, University of Cape Town, Private Bag X3, Rondebosch 7701, South Africa\\
$^{5}$National Institute for Theoretical and Computational Sciences (NITheCS), KwaZulu-Natal, South Africa\\
$^{6}$Department of Physics, CCIS 4-181, University of Alberta, Edmonton, AB, T6G 2E1, Canada\\
$^{7}$Center for Astro, Particle and Planetary Physics (CAP$^3$), New York University Abu Dhabi, PO Box 129188, Abu Dhabi, UAE\\
$^{8}$INAF, Istituto di Astrofisica Spaziale e Fisica Cosmica, Via U. La Malfa 153, I-90146 Palermo, Italy\\
$^{9}$Anton Pannekoek Institute for Astronomy, University of Amsterdam, Science Park 904, 1098 XH Amsterdam, The Netherlands\\
$^{10}$INAF–Osservatorio Astronomico di Brera, via E. Bianchi 46, 23807 Merate (LC), Italy\\
$^{11}$International Centre for Radio Astronomy Research, Curtin University, GPO Box U1987, Perth, WA 6845, Australia\\
}

\date{Accepted XXX. Received YYY; in original form ZZZ}

\pubyear{2021}

\begin{document}
\label{firstpage}
\pagerange{\pageref{firstpage}--\pageref{lastpage}}
\maketitle

\begin{abstract}
Vela X-1 is a runaway X-ray binary system hosting a massive donor star, whose strong stellar wind creates a bow shock as it interacts with the interstellar medium. This bow shock has previously been detected in H$\alpha$ and IR, but, similar to all but one bow shock from a massive runaway star (BD+43$^{\rm o}$3654), has escaped detection in other wavebands. We report on the discovery of $1.3$ GHz radio emission from the Vela X-1 bow shock with the MeerKAT telescope. The MeerKAT observations reveal how the radio emission closely traces the H$\alpha$ line emission, both in the bow shock and in the larger-scale diffuse structures known from existing H$\alpha$ surveys. The Vela X-1 bow shock is the first stellar-wind-driven radio bow shock detected around an X-ray binary. In the absence of a radio spectral index measurement, we explore other avenues to constrain the radio emission mechanism. We find that thermal/free-free emission can account for the radio and H$\alpha$ properties, for a combination of electron temperature and density consistent with earlier estimates of ISM density and the shock enhancement. In this explanation, the presence of a local ISM over-density is essential for the detection of radio emission. Alternatively, we consider a non-thermal/synchrotron scenario, evaluating the magnetic field and broad-band spectrum of the shock. However, we find that exceptionally high fractions ($\gtrsim 13$\%) of the kinetic wind power would need to be injected into the relativistic electron population to explain the radio emission. Assuming lower fractions implies a hybrid scenario, dominated by free-free radio emission. Finally, we speculate about the detectability of radio bow shocks and whether it requires exceptional ISM or stellar wind properties.
\end{abstract}

\begin{keywords}
X-rays: binaries -- stars: early-type -- stars: individual: HD 77581 -- radio continuum: general -- shock waves
\end{keywords}



\section{Introduction}

Astrophysical bow shocks are extended shock nebulas observed around a wide range of different objects. For instance, such structures have been observed in pulsars, indicating shocks between the pulsar wind and the ISM \citep[e.g.][]{cordes1993,gaensler2000,stappers2003}, interacting binary systems (e.g. the fast-moving low-mass X-ray binary SAX J1712.6-3739; \citealt{wiersema2009}; or the nova-like cataclysmic variable, V341 Ara; \citealt{castro2021}). Bow shocks can be created by outflows from the object in question, such as the jet launched by the X-ray binary Cyg X-1 \citep{russell2007}. Commonly, however, bow shocks are created when the object itself moves through the interstellar medium (ISM) at a velocity exceeding the local sound speed. The latter is the case for massive runaway stars, where their strong stellar wind sweeps up interstellar material, creating a forward shock moving at the star’s velocity through the ISM, and a reverse shock at the (higher) stellar wind velocity. The arc-like morphology of the bow shock is determined both by stellar properties, such as mass-loss rate, terminal wind velocity, and stellar bulk motion, as well as the local ISM properties \citep{comeron1998,meyer2016}.

Stellar bow shocks are most often detected via shock-heated dust emission at IR wavelengths: after the first systematic IR searches in IRAS data \citep[see e.g.][]{noriega1997}, subsequent catalogues using \textit{WISE} \citep[the Extensive stellar BOw Shock Survey, or E-BOSS; ][]{peri2012,peri2015} and \textit{Spitzer} \citep{kobulnicky2016} have built up a sample of more than 700 IR stellar bow shocks. Optical line emission, such as H$\alpha$ \citep[e.g.][]{kaper1997,gvaramadze2018} or [O III] $\lambda 5007$ \citep{gull1979}, has also been detected in a much smaller subset of bow shocks; a complicating factor in detecting H$\alpha$ line emission may be that it can be screened by the presence of HII regions surrounding the runaway star \citep{brown2005,meyer2016}.

Bow shock emission is expected in bands beyond the optical and IR. In the shocks between the stellar wind and the ISM, charged particles can be accelerated to relativistic energies via the Diffuse Shock Acceleration mechanism \citep[e.g.][and references therein for reviews]{bell1978a,bell1978b,drury1983,matthews2020}. The resulting relativistic electron population can emit non-thermally across the entire electromagnetic spectrum \citep{delvalle2012,delvalle2018,delpalacio2018}, primarily via synchrotron processes (expected to dominate at radio wavelengths) and inverse Compton scattering of IR and stellar photons (dominant at and/or above X-ray wavelengths). In addition, thermal free-free emission (Bremsstrahlung), from the same electron population that causes the optical line emission, is expected at long wavelengths. The relative contribution of these non-thermal and thermal processes depends on the energetics of the stellar winds, the electron density and temperature in the shock, as well as the shock acceleration efficiency of electrons.

Despite this set of emission processes, bow shock detections outside the optical and IR bands remain exceptionally rare around massive runaway stars. At radio wavelengths, only a single such bow shock has been detected: BD+43$^{\rm o}$3654 was detected with the Karl G. Jansky Very Large Array \citep[VLA;][]{benaglia2010,benaglia2021} and the Giant Metrewave Radio Telescope \citep[GMRT;][]{brookes2016} with a spatially variable and generally steep spectrum (defined as $\alpha>0$ where $S_\nu \propto \nu^{-\alpha}$), which has been interpreted as non-thermal synchrotron emission \citep{benaglia2010}. Despite further searches in other IR-identified bow shocks, no other radio counterparts have yet been identified in surveys or pointed observations \citep[][the latter based on private communication with C. Peri]{rangelov2019,benaglia2021}. Similarly, the expected non-thermal X-ray or $\gamma$-ray emission has not been unambiguously detected, despite several searches \citep{schulz2014,toala2016,toala2017,debecker2017,hess2018}\footnote{\citet{lopezsantiago2012} reported an X-ray detection of the bow shock of AE Aurigae with XMM-Newton. However, a later re-analysis by \citet{toala2017} did not confirm this result.}. The best candidates of possible high-energy bow shock counterparts are reported by \citet{sanchezayaso2018}, who associate two unidentified Fermi sources with known bow shocks. 

In this work, we present the MeerKAT radio detection of the Vela X-1 bow shock. Vela X-1 is a high-mass X-ray binary (HMXB) system, consisting of an accreting neutron star and the supergiant donor HD 77581 separated at $\sim$1.7 donor-star radii ($53.4$ $\rm R_{\odot}$) in a tight $\sim$8.96-day orbit. Vela X-1 is a runaway system, travelling at a bulk velocity of $\sim54.3\pm0.6$ km/s 
\citep[][see Section \ref{sec:thermal}]{gvaramadze2018}. The supergiant donor star launches a strong stellar wind, whose mass loss rate and terminal velocity have been estimated in numerous earlier studies to lie in the range $\sim 5\times10^{-7}$ to $5\times10^{-6}$ $\rm M_{\odot}$ per year and $\sim 500$ to $1700$ km/s, respectively \citep[see e.g.][and references therein for an in-depth review of this HMXB]{kretschmar2021}. Similarly, many literature estimates of its distance exist: here, we adopt the Gaia DR2 distance of $1.99_{-0.11}^{+0.13}$ kpc derived by \citet{kretschmar2021} and refer to that work for further discussion. 

The Vela X-1 stellar wind creates a bow shock, which was first discovered in narrow-band H$\alpha$ imaging by \citet{kaper1997}. To date, Vela X-1 remains one of only two HMXBs with a bow shock \citep[with 4U 1907+09;][]{gvaramadze2011}. The bow shock of Vela X-1 was later detected at IR wavelengths as well \citep{peri2015,maizapellaniz2018}, and its morphology suggests that Vela X-1 may have originated in the Vel OB1 association several Myr ago \citep{kaper1997}. Using the SuperCOSMOS H$\alpha$ Survey\footnote{\href{http://www-wfau.roe.ac.uk/sss/halpha/}{http://www-wfau.roe.ac.uk/sss/halpha/}}, \citet{gvaramadze2018} reveal the presence of large-scale diffuse structure in the wake of the binary and its bow shock; their detailed simulations indicate that Vela X-1 interacts with a local ISM overdensity, where an ISM density of approximately three times that of the ambient density can reproduce the observed large-scale morphology and H$\alpha$ surface brightness. 

Here, we report the serendipitous detection of radio emission from the Vela X-1 bow shock in MeerKAT observations targeting the HMXB itself. In Section 2, we present the MeerKAT data and analysis methods, while the results are shown in Section 3. Then, in Section 4, we perform initial analytic calculations to assess the origin and nature of the radio emission, comparing a non-thermal and thermal scenario. 

\section{Observations and data analysis}

\begin{figure*}
 \includegraphics[width=\textwidth]{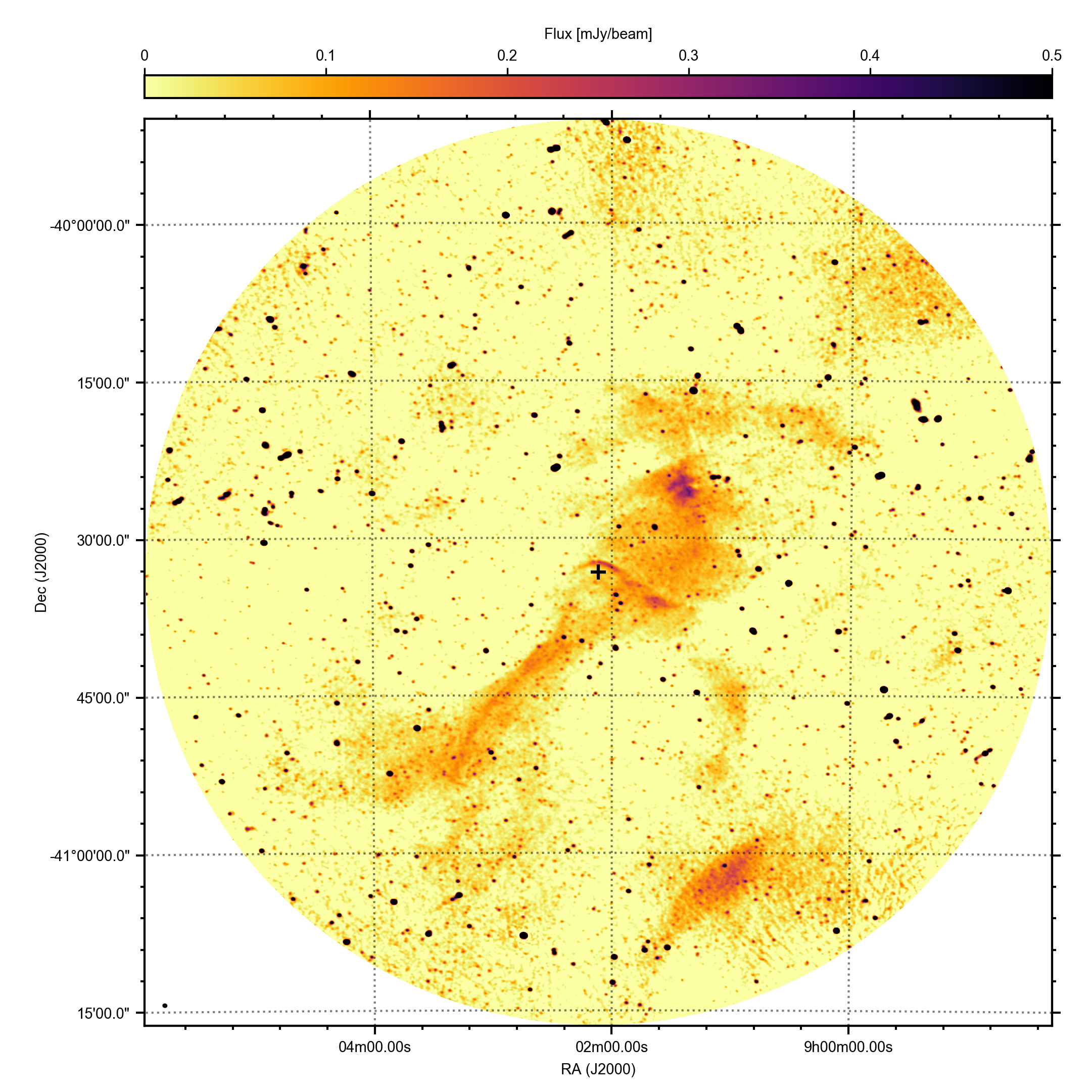}
 \caption{Radio continuum emission of the full MeerKAT field of view of Vela X-1, shown by the cross, and its surroundings, created by combining three observing runs. The combined exposure time is $90$ minutes, yielding a $40$ $\mu$Jy RMS sensitivity. The observations were performed at L-band (1.3 GHz), and reveal both the Vela X-1 bow shock as well as several other large-scale diffuse structures in radio for the first time. The beam size is shown in the bottom left of the image.}
 \label{fig:fullfield}
\end{figure*}

We observed the field around the HMXB Vela X-1 as part of the ThunderKAT Large Survey Project with MeerKAT \citep{fender2017}, which performs radio observations of active, Southern X-ray binaries, cataclysmic variables, supernovae, and gamma-ray bursts. Two of the three MeerKAT observations included in this work were part of a larger coordinated multi-wavelength campaign of the binary; the results of that study, including the MeerKAT results on Vela X-1 itself, will be reported elsewhere (Van den Eijnden et al., \textit{in prep.}). In this paper, we analyse only the Stokes I data obtained in these observations.

Vela X-1 (J2000 09$^{\rm h}$02$^{\rm m}$06.86$^{\rm s}$ $-$40$^{\circ}$33$^{'}$16.9$^{''}$) was observed with the MeerKAT telescope on 2020-09-25, 2020-09-27, and 2020-10-11 (capture block IDs 1600995961, 1601168939, and
1602387062 respectively). The telescope's L band (856 -- 1712 MHz; hereafter reported at the central frequency of 1.3 GHz) receivers were used, with the correlator configured to deliver 32,768 channels and 8 second integration time per visibility point. For the three observations there were 59, 61, and 60 antennas used in the array.

Vela X-1 was observed for 30 minutes in each of the three runs, book-ended by two 2-minute scans of the nearby secondary calibrator source J0825$-$5010. The standard primary calibrator source J0408-6545 was also observed for 5 minutes at the start of each run. 

For each of the three observations, reference calibration was performed using the {\sc casa} package \citep{mcmullin2007}. Bandpass, delay and flux-scale corrections were derived from the scans of the primary, and complex gain and delay corrections obtained from the scans of the secondary, following all flagging of the data. These corrections were applied to the target data, which was then split and flagged using the {\sc tricolour}\footnote{\url{https://github.com/ska-sa/tricolour/}} package. The target data were imaged using {\sc wsclean} \citep{offringa2014}, reimaged following the construction of a deconvolution mask, and then self-calibrated using the {\sc cubical} package \citep{kenyon2018} to solve for phase and delay corrections for every 32 seconds of data. 

The self-calibrated data for all three observations were then jointly imaged using {\sc ddfacet} \citep{tasse2018}. Direction-dependent gain corrections were derived using {\sc killms} \citep{smirnov2018}, with the sky partitioned into 15 directions governed by the distribution of bright, compact sources in the field. The data were then re-imaged using {\sc ddfacet}, applying the directional gain corrections in the process. The final wide-field image presented in Figure \ref{fig:fullfield} has been primary beam corrected using the {\sc katbeam} package\footnote{\url{https://github.com/ska-sa/katbeam}}, blanking the map beyond the nominal 30\% level. Further details on the data reduction process together with the relevant calibration and imaging parameters can be found online\footnote{\url{https://github.com/IanHeywood/oxkat}} \citep{heywood2020}.

\section{Results}

\subsection{MeerKAT observations}

\begin{figure}
 \includegraphics[width=\columnwidth]{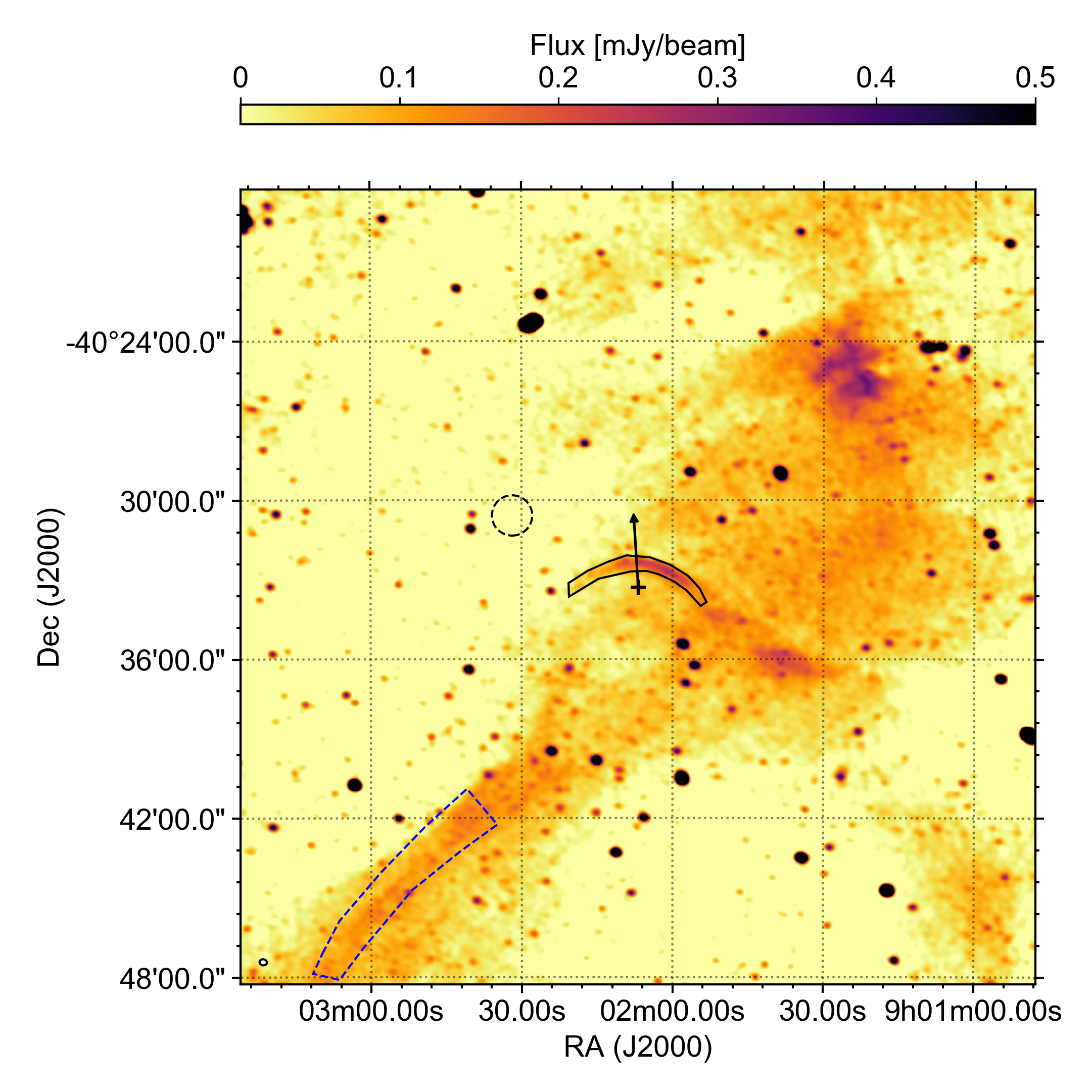}
 \caption{Inset of the Vela X-1 field shown in Figure 1, extending 0.25 degree out from the HMXB (indicated by the cross) in RA and Dec. The $12$ arcsecond circular beam size is shown in the bottom left. The bow shock source region and background region are shown as the black lined and dashed regions; the blue dashed region indicates the comparison overdense region discussed in Section \ref{sec:thermal}. The arrow shows the direction of motion of Vela X-1 relative to its surroundings.}
 \label{fig:zoom}
\end{figure}


\begin{figure}
    \includegraphics[width=\columnwidth, frame]{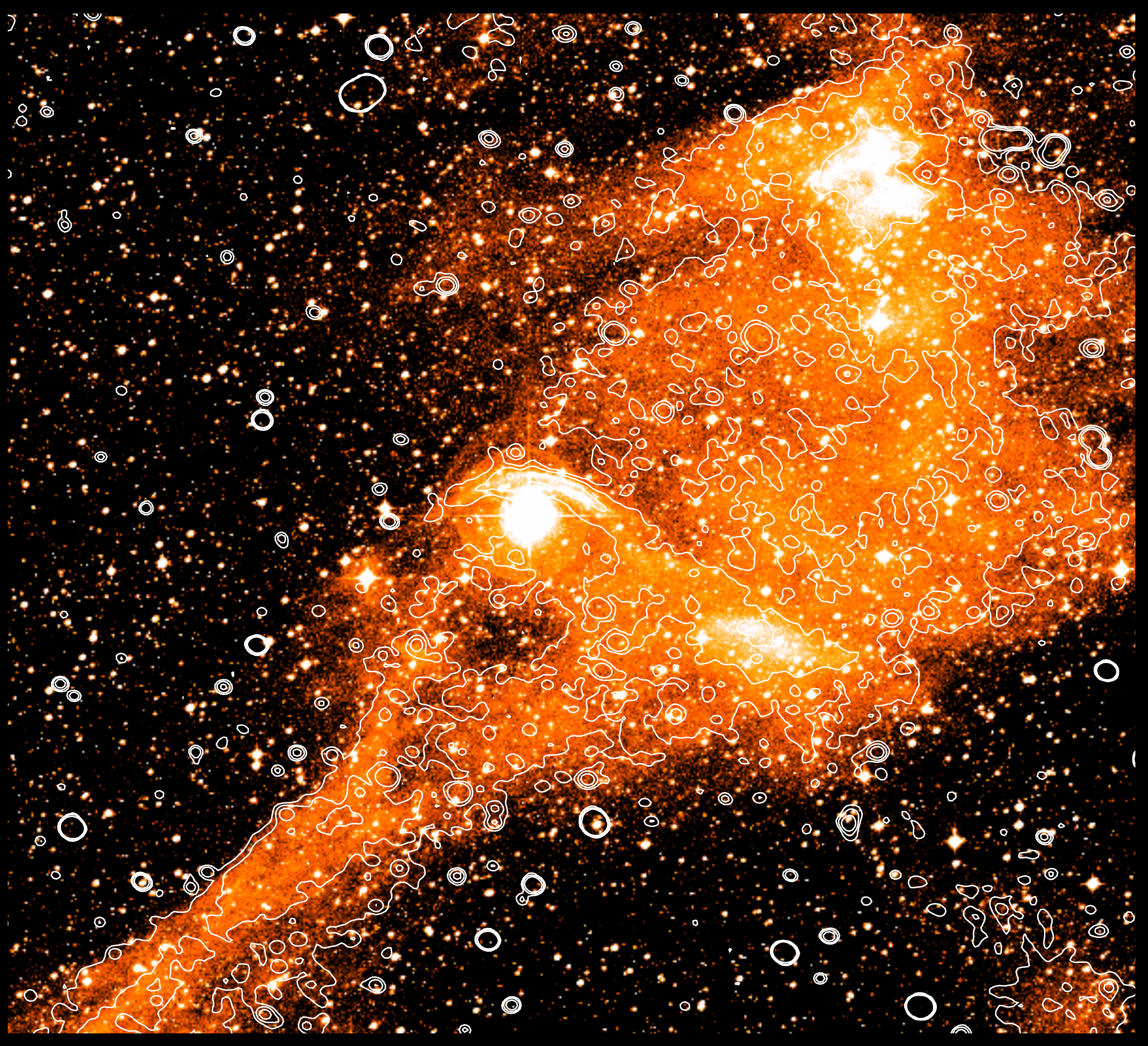}
    \caption{the SuperCOSMOS H$\alpha$ image, overlaid with the contours from the MeerKAT L-band image. The radio contours are shown at levels of $50$, $100$, and $200$ $\mu$Jy, calculated after smoothing the image by a factor four. These levels are chosen to highlight the diffuse structures in the radio image. The large-scale H$\alpha$ and radio structures appear to trace each other closely throughout the field.}
   \label{fig:contours}
\end{figure}

\begin{figure}
    \includegraphics[width=\columnwidth]{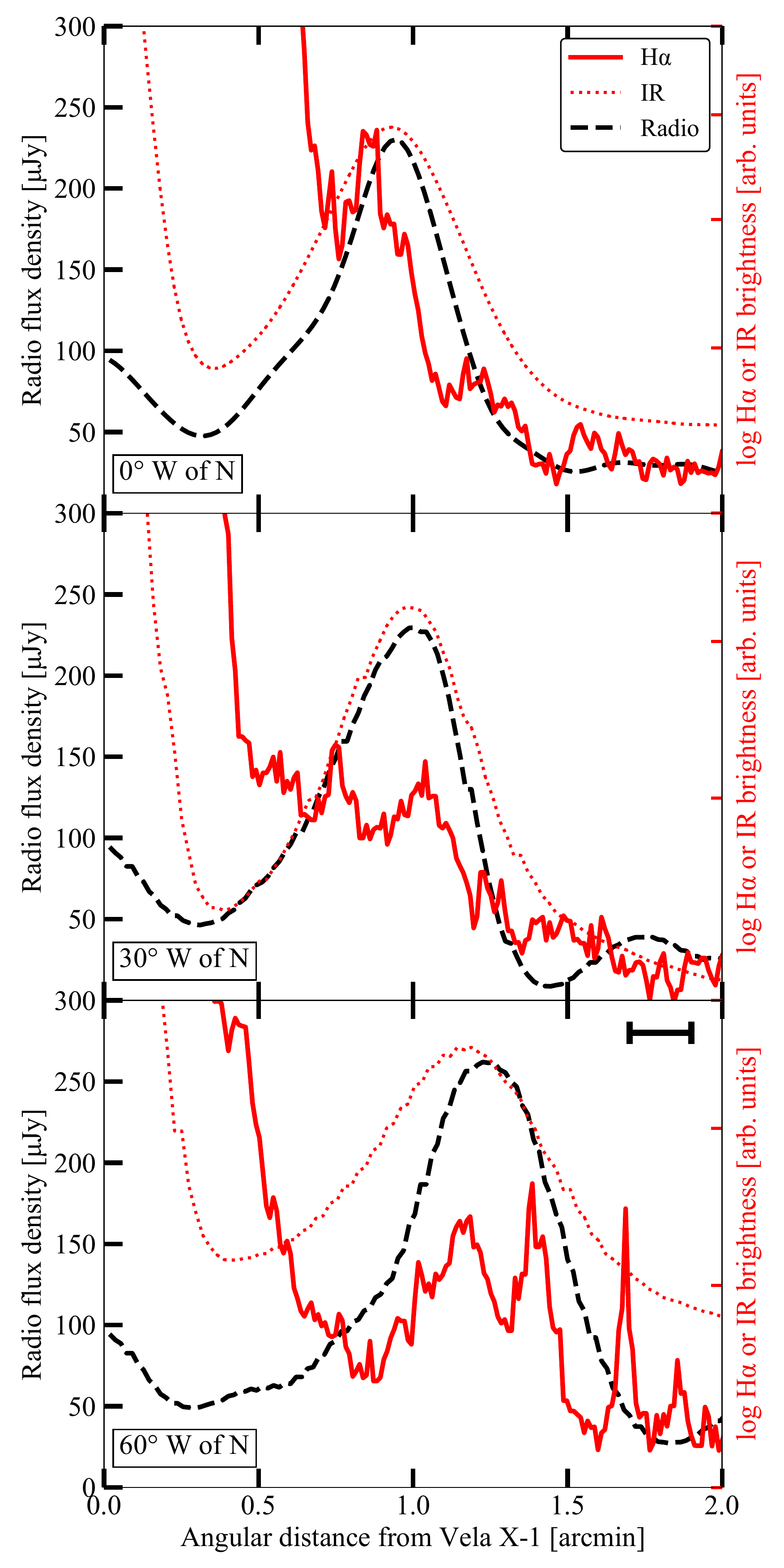}
    \caption{The radial flux density or brightness profiles at three directions from Vela X-1, namely 0, 30, and 60 degrees West of North. The black dashed line shows the radio profile, while the red line and dotted line show the logarithmic H$\alpha$ and IR profiles, with arbitrary units, respectively.The black bar in the top right of the bottom panel, shows the beam size of the MeerKAT observation. The difference in resolution between bands is clearly visible.}
   \label{fig:lines}
\end{figure}

In Figure \ref{fig:fullfield}, we show the combined, full-field 1.3-GHz MeerKAT image created from the three observations, extending $\sim$0.72 degree out from the central pointing towards the HMXB. Multiple large-scale, diffuse structures are visible, including the bow shock immediately above the cross indicating Vela X-1. Similarly, an extended, diagonal structure is observed, which Vela X-1 appears to have crossed through. Both structures, as well as the other extended radio sources in the image, were already known from H$\alpha$ images of this field \citep[i.e.][using SuperCOSMOS data]{gvaramadze2018}. In Figure 2, we show an inset containing the central $0.5\times0.5$ degree square of the field, providing a more detailed view of the bow shock region.

Furthermore, in Figure \ref{fig:contours}, we overlay radio contours on the SuperCOSMOS image, indicating the remarkable similarity between the observed morphologies in both H$\alpha$ and L-band imaging. This morphological similarity, combined with their relatively similar order-of-magnitude flux densities, suggests a possible connection between the emission mechanisms in the bow shock and the surrounding structures. Such a connection could be explained via thermal processes that are combined with shock-enhanced electron densities in the bow shock, which we will investigate in detail in the next section. While high-spatial-resolution H$\alpha$ images reveal filamentary structures in the bow shock \citep[e.g.][]{kaper1997}, our L-band MeerKAT image does not reveal similar features. However, this can be attributed to the larger, 12" beam size of the MeerKAT observations, which is similar to the width of the H$\alpha$ filaments and larger than the separation between them.

A more detailed comparison of the bow shock profile is shown in Figure \ref{fig:lines}, where we show the radio flux density from MeerKAT, H$\alpha$ brightness from the SuperCOSMOS Survey, and IR brightness in the W3 band of \textit{WISE}\footnote{Obtained from \url{https://irsa.ipac.caltech.edu/applications/wise/}.} along three lines from Vela X-1, angled at 0, 30, and 60 degrees West of North. These comparisons show how all three bands, to their resolution, show compatible bow shock profiles. The H$\alpha$ band, dominated by the stellar PSF at small distances, shows a single peaked profile at the bow shock apex, but a double peaked, filamentary profile for the other directions (note that the third peak in the bottom panel, at $\sim 1.6$--$1.7$ arcmin, is due to a star). The lower resolution of the IR and radio images (the latter shown by the black bar in the bottom panel), smear out any sub structure that may be present, leaving a broad and smoothly peaked profile. 

To measure the radio flux densities of the bow shock, we construct a custom region (black polygon) in ds9 \citep{joye2003}, as shown in Figure 2. This region was constructed manually, as the combination of increasing flux along shock in the West-ward direction, with the surrounding over-dense region on its West-end, prevents using a single flux contour level to define the entire shock. For angles West of North along the shock, the region's edge is defined where the flux density drops below approximately 100 $\mu$Jy, extending West-wards up to point where the H$\alpha$ shock morphology can be identified. On the fainter East side of the shock, un-surrounded by diffuse emission, the shock edge is defined where it drops approximately below the RMS defined below. We discuss the effects of this region definition in Section \ref{sec:disc}.

The peak radio flux density of the entire bow shock is $S_{\nu, \rm max} = 270 \pm 40$ $\mu$Jy/bm, where the error equals the RMS in the nearby region (dashed circle in Figure \ref{fig:zoom}) devoid of point sources and strong diffuse emission. We also apply the \textsc{radioflux}\footnote{\href{https://github.com/mhardcastle/radioflux}{https://github.com/mhardcastle/radioflux}} script to measure the integrated radio flux across the entire bow shock, finding $S_{\nu, \rm total} = 5.27 \pm 0.07$ mJy, corresponding to a mean flux density of $S_{\nu, \rm mean} = 130 \pm 40$ $\mu$Jy/bm. For our later modelling (Section 4), we also note that the projected bow shock area assumed in these calculations is 10100 arcsec$^2$. While the bow shock structure can also be identified in the sub-bands that are not strongly affected by RFI, we do not calculate an in-band spectral index; even when averaging across the entire shock, the S/N is not sufficient for a reliable measurement \footnote{Moreover, the analysis of the radio bow shock of BD+43 3652 shows the spectral index can vary strongly with position in the shock \citep{benaglia2010,benaglia2021,brookes2016}.}.

\subsection{Existing ASKAP and \textit{Chandra} data}

\begin{figure*}
 \includegraphics[width=\textwidth]{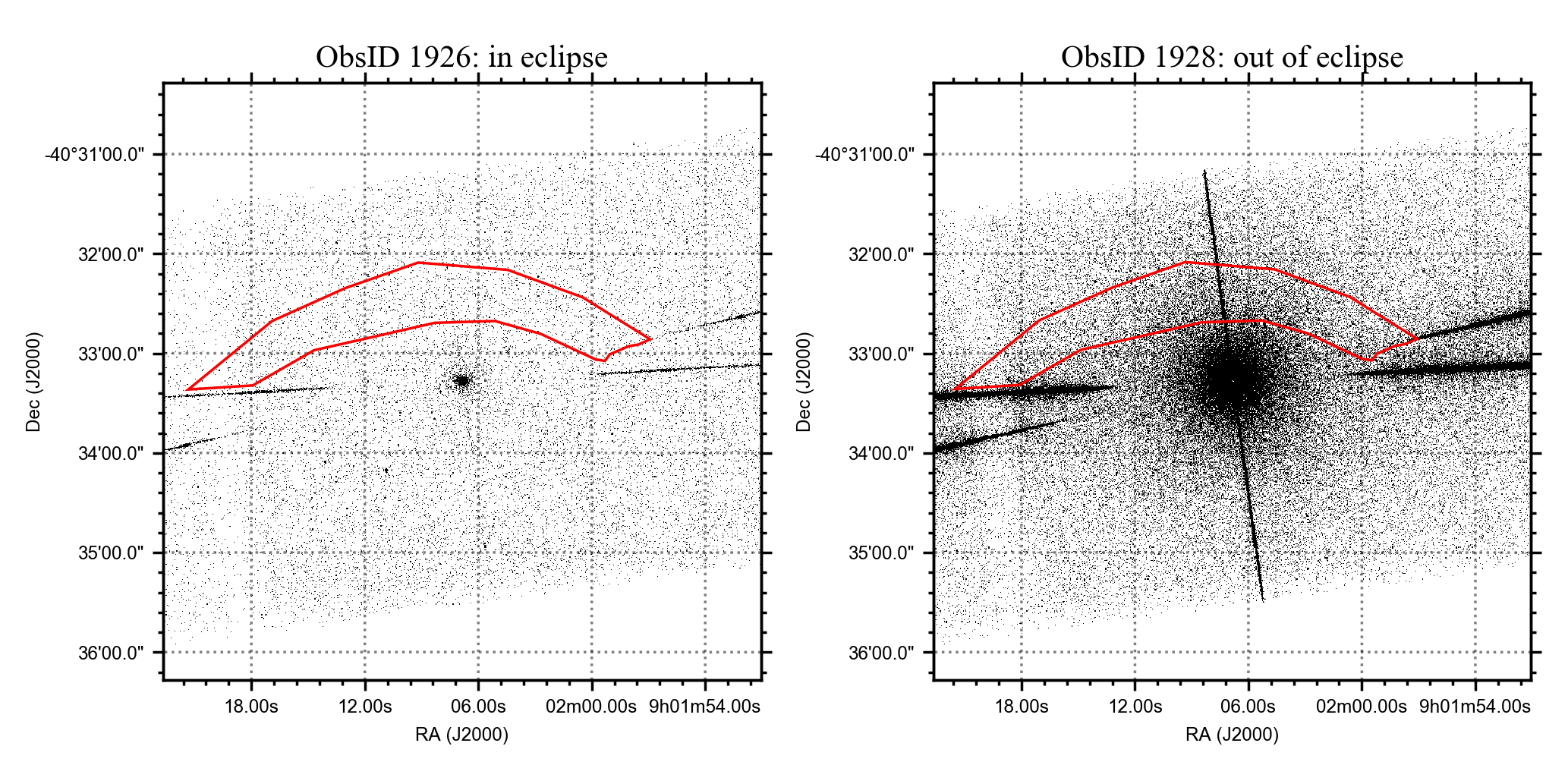}
 \caption{The 0$^{\rm th}$ order S3 chip data taken by \textit{Chandra} during eclipse (ObsID 1926; left) and out of eclipse (ObsID 1928, right). The red region indicates the same bow shock region as in Figure 2, adjusted slightly to avoid the X-shaped grating arms. Despite the lack of counts from the HMXB leaking into the shock region during eclipse, no bow shock counterpart can be identified.}
 \label{fig:chandra}
\end{figure*}

To further constrain the spectral shape, we therefore instead opt to use observations with different observatories, at different frequencies. First, we consulted the data release of the Rapid ASKAP Continuum Survey \citep[RACS;][]{mcconnell2020}, which provides complete sky coverage at declinations below +41$^{\rm o}$ at a spatial resolution similar to MeerKAT ($\sim$ 15 arcsec). All images in the first data release are taken at a central frequency of 887.5 MHz with a 288 MHz bandwidth, just below (but overlapping with) the MeerKAT L-band. The bow shock location is covered by three pointings -- 0911-37A, 0840-37A, and 0855-43A -- with median RMS sensitivities of 285, 292, and 222 $\mu$Jy. No hints of the bow shock are visible in any of the three images. Combining the latter non-detection with the mean MeerKAT L-band flux density of $130\pm40$ $\mu$Jy/bm, implies an upper limit on the spectral index of $\alpha\geq1.4$ (where $S_\nu \propto \nu^{-\alpha}$) -- which is not constraining for any realistic emission process in the bow shock. Furthermore, the different array configurations of MeerKAT and ASKAP, with the former’s core-heavy distribution especially suitable for the detection of extended emission, further complicates the comparison between the two radio observatories.

If the observed radio emission originates from non-thermal synchrotron processes, it indicates the presence of a population of relativistic electrons accelerated in the shock. As discussed in more detail in Section 4, such an electron population can generate X-ray emission either as the high-energy part of its synchrotron spectrum, or inverse-Compton scattering of dust and stellar photons. Whether the former contributes in the X-ray band depends on physically interesting parameters such as the maximum electron energy and the magnetic field. Hence, we also investigated the X-ray properties of the bow shock.

As Vela X-1 is a persistently-accreting HMXB, its X-ray luminosity (typically $\sim 10^{36}$ erg/s) will overpower any diffuse emission from the bow shock. Given its X-ray luminosity and  relatively small (less than one arcminute) angular distance between the binary and the shock, the point-spread function of Vela X-1 overlaps with the shock for all X-ray observatories. However, Vela X-1 is viewed almost edge on and consistently shows eclipses between orbital phases $\sim$0.9 and $\sim$0.1 \citep{falanga2015}, when the \textit{Chandra} count rate drops by two orders of magnitude and most of the X-ray flux is concentrated in a handful of narrow emission lines \citep{watanabe2006}. \textit{Chandra} observed the field of Vela X-1 twice during eclipses (ObsIDs 102 and 1926, $\sim 28$ and $\sim 83$ ks exposures, respectively), both times with the High-Energy Transmission Grating (HETG) employed. While this instrumental setup is optimized for high-resolution X-ray spectroscopy, the observation provides an image of the observed field from the 0$^{\rm th}$ order data collected on chip S3. Therefore, we downloaded both observations, reprocessed the data using the task \textsc{chandra\_repro} in \textsc{ciao} version 4.13 \citep{fruscione2006}\footnote{\href{https://cxc.cfa.harvard.edu/ciao/}{https://cxc.cfa.harvard.edu/ciao/}}, and finally generated an image by using \textsc{dmcopy}, selecting only photons between 0.5-10 keV and screening cosmic-ray-induced events below 2 keV. 

In Figure \ref{fig:chandra}, we show the resulting 0$^{\rm th}$-order image of ObsID 1926 on the left, alongside ObsID 1928, which was taken outside eclipse, for comparison on the right. During eclipse, despite the faintness of Vela X-1, the HMXB is detected as the point source, and faint gratings arms are visible in their characteristic X-shaped pattern. The red region in both panels shows the MeerKAT L-band bow shock region, adjusted slightly to avoid the arms. While the binary PSF only leaks into the shock region outside of eclipse, no evidence for an X-ray bow shock can be seen in the eclipse observation. For the two eclipse observations 102 and 1926, we measure background count rates in the full shock region $(2.72\pm0.05)\times10^{-2}$ and $(3.2\pm0.1)\times10^{-2}$ cts/s. These rates are slightly lower than the approximate background rate for the entire chip, scaled by the surface of the region, confirming the above conclusion that no significant bow shock X-ray emission is detected. To convert these count rates to a flux upper limit, we use the S3 background spectrum\footnote{\href{https://cxc.harvard.edu/contrib/maxim/bg/index.html}{https://cxc.harvard.edu/contrib/maxim/bg/index.html}}, finding that the former count rate corresponds to a flux of approximately $7\times10^{-9}$ Jy at 1 keV. Therefore, we set a 3-$\sigma$ upper limit to the integrated bow shock X-ray flux at 1 keV of $2\times10^{-8}$ Jy. 

\section{Discussion}
\label{sec:disc}

\begin{table*}
\caption{The relevant parameters in our thermal and non-thermal scenario calculations, with their assumed values and references.}
\label{tab:input}
\begin{tabular}{llll}
\hline
Parameter & Quantity & Value & Reference \\ \hline
$S_{\nu, \rm total}$ & Total radio flux density & $5.27 \pm 0.07$ mJy & \textit{This work} \\ %
$S_{\nu, \rm max}$ & Maximum radio flux density & $270 \pm 40$ $\mu$Jy/bm & \textit{This work} \\ %
$S_{\nu, \rm mean}$ & Mean radio flux density & $130 \pm 40$ $\mu$Jy/bm & \textit{This work} \\ %
$\nu_{\rm obs}$ & MeerKAT central frequency & $1.3$ GHz & \textit{This work} \\ %
$\theta_b$ & MeerKAT beam size (circular) & $12$ arcsec & \textit{This work} \\ %
$S_{\rm H\alpha}$ & H$\alpha$ surface brightness & $1.0^{+1.0}_{-0.5}\times10^{-15} $ erg s$^{-1}$ cm$^{-2}$ arcsec$^{-2}$ & \citet{kaper1997}\\ %
& & $2.5\times10^{-16}$ erg s$^{-1}$ cm$^{-2}$ arcsec$^{-2}$ & \citet{gvaramadze2018} \\ %
$F_{1\text{ keV}}$ & X-ray flux density at 1 keV & $<2\times10^{-8}$ Jy & \textit{This work} \\ \hline
$R_0$ & Standoff distance & $0.57$ pc & \citet{gvaramadze2018} \\ %
$\Delta$ & Bow shock width & $0.31$ pc & \textit{This work} \\ %
$A$ & Bow shock surface & $10100$ arcsec$^2$ & \textit{This work} \\ %
$D$ & Distance & $1.99$ kpc & \citet{kretschmar2021} \\ %
$V_{\rm bowshock}$ & Bow shock volume & $0.28$ pc$^3$ & \textit{This work} \\ %
$\eta_{\rm vol}$ & Volume factor & $0.36$ & \textit{This work} \\ \hline
$\dot{M}_{\rm wind}$ & Wind mass loss rate & $10^{-6}$ $\rm M_{\odot}$/yr & \citet{grinberg2017}* \\ %
$v_\infty$ & Wind terminal velocity & $700$ km/s & \citet{grinberg2017}* \\ %
$v_*$ & XRB velocity & $54.25$ km/s & \citet{gvaramadze2018} \\ %
\hline
\end{tabular}\\
\textit{*See also \citet{kretschmar2021} for an extensive discussion regarding these parameters.}
\end{table*}

In this work, we report the 1.3-GHz radio discovery of the bow shock of the HMXB Vela X-1 with the MeerKAT telescope. We do not detect a counterpart at either lower radio frequencies (using RACS) or X-ray energies (\textit{Chandra}). Therefore, we lack the most direct observable that can help distinguish between a thermal and non-thermal origin of the radio emission: a spectral index measurement\footnote{Although we note that while a steep spectral index can be regarded as evidence for a non-thermal scenario, a flatter spectral index is consistent with both options. Therefore, a spectral measurement alone also does not necessarily solve this issue.}. Instead, in this discussion, we will perform simple, analytical calculations to compare these two scenarios, and investigate their compatibility with the energetics of the system and the constraints from other wavebands. To increase the reproducibility of our results, these calculations can be repeated in the Jupyter notebook accompanying this paper (see the Data Availability statement). We stress, however, that these two processes are not mutually exclusive. As we discuss briefly at the end of Section \ref{sec:nonthermal}, both processes are likely to co-exist and contribute to the total radio emission.

\subsection{General considerations}
\label{sec:input}

Before evaluating the two scenarios mentioned above, we briefly discuss some general considerations for these calculations. Firstly, we should consider whether the system is in a steady state, given its prior interactions with the over-dense region \citep{gvaramadze2018}. Given the separation of $\sim 5$ arcmin between the current position of Vela X-1 and this over-density, tracing back Vela X-1's direction of motion (shown by the arrow in Figure \ref{fig:zoom}), and its space velocity of $\sim 54.3$ km/s and $1.99$ kpc distance, this interaction likely took place $\sim 5.5\times10^4$ yr ago. Simulations of bow shock formation in, for instance, Betelgeuse show how a steady state is achieved at similar time scales, even though no bow shock is present at the start of the simulations \citep{mohamed2012}. A similar conclusion follows from comparing the standoff distance of the bow shock to its distance from Vela X-1 at a 90 degree angle from the apex: in radio, this ratio $R(0)/R(90) = 0.50\pm0.12$, consistent with the expected value of $0.57$ for the shape of a steady state bow shock \citep{wilkin1996}. Based on these two lines of reasoning, we will thus assume the bow shock is in a steady state.

For our calculations, we will assume that the bow shock is homogeneous throughout, which enables us to perform simple analytic calculations and estimates. In reality, the radio image, as well as the filamentary H$\alpha$ image, show that this assumption does not hold. However, we prioritize performing analytic calculations at this stage, and will leave the spatially resolved analyses to future works using simulations beyond the scope of this work. In addition, the radio flux density varies slightly on the scale of the MeerKAT beam: the peak and average flux density differ by a factor $\sim 2$. We expect that our physical estimates will trace the variations across the shock to order of magnitude, on the scales of the beam size (roughly $4$\% of the bow shock region in Figure \ref{fig:zoom}). In future studies, a spatially resolved analysis would be interesting to understand this spatial flux density variations and its relation to the emission's origin. Similarly, future studies of the spectral shape may probe whether the bow shock region defined for this work, fully traces the region where the shock's radio emission is dominant.

For geometrical purposes, we will follow the common assumption that the bow shock has a depth that is of the order of its width, $\Delta$. In the radio image, we measure $\Delta \approx 35$ arcsec, comparable to the standoff distance $R_0 \approx 65$ arcsec. Given the surface area of the bow shock region ($10100$ arcsec$^2$) and distance to Vela X-1, this translates to an approximate bow shock volume of $V_{\rm bowshock} = 0.28$ pc$^3$. The bow shock volume factor, i.e. the fraction of the wind power passing through the bow shock, can be estimated as $\eta_{\rm vol} \equiv V_{\rm bowshock}/(4\pi R_0^3/3) \approx 0.36$. If we instead use the approach by \citet{debecker2017}, which compares surface fluxes, we find $\eta_{\rm vol} \approx 0.15$. This difference arises from the non-circular shape of the bow shock arc. In our following calculation, using a higher $\eta_{\rm vol}$ implies lower injection efficiencies (Section \ref{sec:nonthermal}); therefore our assumption on $\eta_{\rm vol}$ makes those calculations more conservative with respect to the conclusions that we draw. In Table \ref{tab:input}, we list all stellar (wind) and bow shock parameters that are relevant to the equations shown in the main text. Those that show up in the Appendix only, are listed there. Finally, we stress that the literature contains many estimates of $\dot{M}_{\rm wind}$, $v_{\infty}$, and $v_*$; we have opted for either representative or recent values and find that none of our qualitative conclusions depend on these exact choices.

\begin{figure*}
 \includegraphics[width=\textwidth]{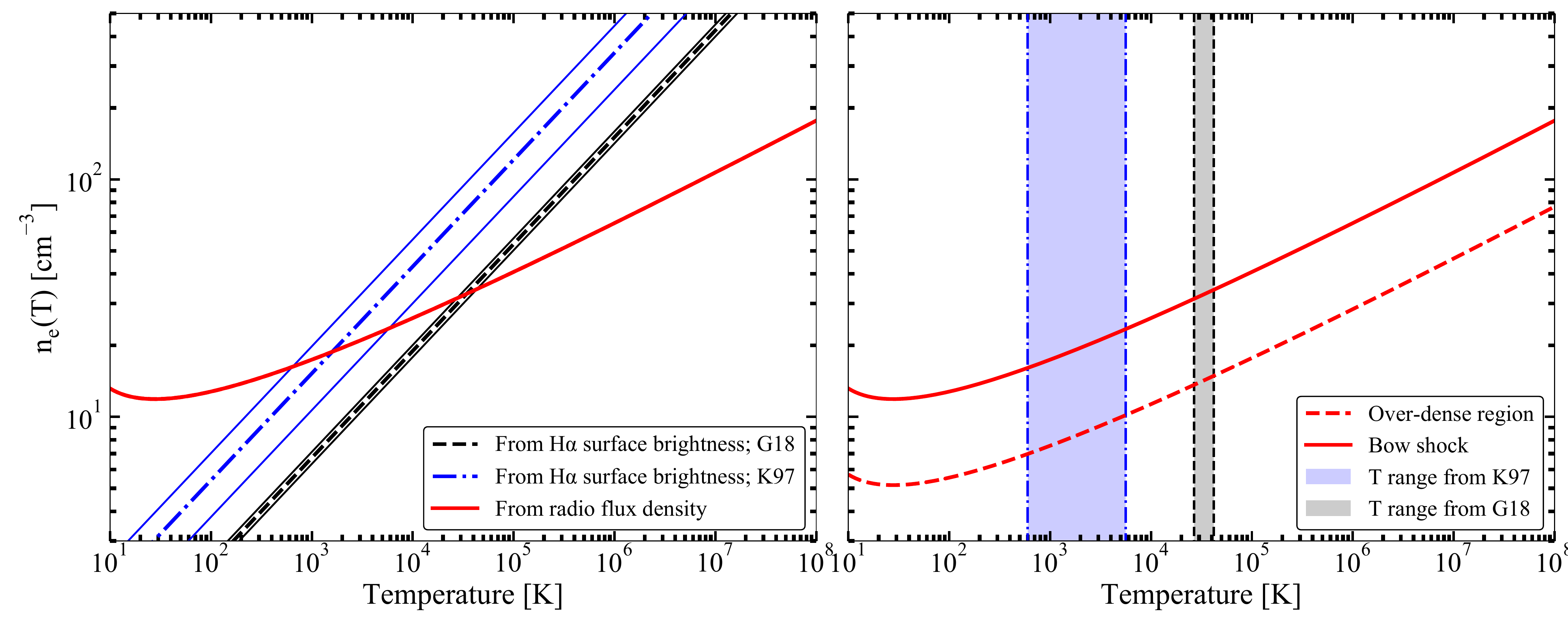}
 \caption{\textit{Left:} the relation between post-shock electron density $n_e$ and temperature $T$ consistent with the observed radio (red line, this work) and H$\alpha$ (blue dash-dotted line, Isaac Newton Telescope, \citet{kaper1997} and black dashed line, SuperCOSMOS H$\alpha$ Survey), \citet{gvaramadze2018}). The intersections between the radio and H$\alpha$ curves indicate solutions that can account for both bands. The narrow blue and black lines indicate the (assumed) uncertainty on the H$\alpha$ surface brightness. \textit{Right:} the same relation between $n_e$ and $T$ for both the bow shock (red line) and the over-dense region (red dashed line) to the bottom-left of Vela X-1 in Figure \ref{fig:contours}. The blue and grey bands indicate the bow shock temperatures where the radio and H$\alpha$ curves intersect in the left panel.}
 \label{fig:ne_from_both}
\end{figure*}

\subsection{The thermal/free-free scenario}
\label{sec:thermal}

Thermal free-free emission (brehmsstrahlung) from a bow shock will depend, fundamentally, on two physical properties of the electron population in the shock: the electron number density $n_e$ and temperature $T$. These parameters will not only set the radio luminosity (and spectrum) of the system, they also determine the surface brightness of H$\alpha$ line emission. As the Vela X-1 bow shock is detected in both H$\alpha$ and radio images, we can combine the wavebands to infer what electron density and temperature are required to explain the observations. 

The free-free emissivity as a function of temperature ($T$) and electron density ($n_e$) is given by \citep{longair2011}:

\begin{equation}
    \kappa_\nu = 6.8\times10^{-38} T^{-1/2} n_e^2 g(\nu, T) e^{\frac{-h\nu}{kT}} \text{ erg cm}^{-3}\text{ s}^{-1}\text{ Hz}^{-1} .
    \label{eq:freefree}
\end{equation}

\noindent In this equation, we have assumed a fully ionized hydrogen gas (i.e. $n_e$ equals the number of protons). At the MeerKAT observing frequency, $h\nu \ll kT$ for any temperature considered in this work. Therefore, we can ignore the final exponent in the above expression. The Gaunt factor $g(\nu, T)$ can be approximated at radio frequencies as \citep{longair2011}:

\begin{equation}
    g(\nu, T) \approx \frac{\sqrt{3}}{2\pi} \left[\ln\left(\frac{128\epsilon_0^2 k^3 T^3}{m_e e^4 \nu^2}\right) - C_{\rm Euler}^2\right] ,
\end{equation}

\noindent where $C_{\rm Euler}$ equals Euler's constant ($\approx 0.577$). The free-free emissivity is related to the observed integrated radio flux (in the optically thin regime, as discussed towards the end of this section) via 

\begin{equation}
    \kappa_\nu = 4\pi D^2 S_{\nu, \rm total}/V_{\rm bowshock} \text{ ,}
    \label{eq:3}
\end{equation}

\noindent where $D$ is the distance to the source, and $V_{\rm bowshock}$ equals the volume of the bow shock. All combined, these equations allow us to determine the relation between electron density and temperature that is consistent with the measured flux density from MeerKAT. 

We can repeat this exercise for the H$\alpha$ line emission, where the surface emissivity is instead given by \citep{gvaramadze2018}:

\begin{equation}
    \kappa_{\rm H\alpha} = 2.85\times10^{-33} \left( \frac{T}{\text{K}} \right)^{-0.9}\left(\frac{n_e}{\text{cm}^{-3}}\right)^2 \text{erg s}^{-1}\text{ cm}^{-3}\text{ arcsec}^{-2} .
    \label{eq:halpha}
\end{equation}

\noindent Integrating this function along the line of sight in the bow shock then yields the surface brightness in H$\alpha$, which can be compared to observed H$\alpha$ maps. As stated, we will assume that the electron density and temperature are homogeneous across the bow shock:

\begin{equation}
    S_{\rm H\alpha} = \int \kappa_{\rm H\alpha}(x) dx \approx \kappa_{\rm H\alpha} \Delta \text{ .}
\end{equation}

\noindent Constraints on the observed H$\alpha$ surface brightness are reported by \citet{kaper1997} and \citet{gvaramadze2018}. The former measure the peak H$\alpha$ intensity to be within a factor 2 of $10^{-15} $ erg s$^{-1}$ cm$^{-2}$ arcsec$^{-2}$, while the latter measure $S_{\rm H\alpha} \approx 2.5\times10^{-16}$ erg s$^{-1}$ cm$^{-2}$ arcsec$^{-2}$. Here, we will use both measurements to infer the relationship between electron density and temperature.

In the left panel of Figure \ref{fig:ne_from_both}, we plot electron density as a function of temperature using the observed MeerKAT flux and both literature $S_{\rm H\alpha}$ constraints. The dashed blue and black lines indicate the factor $2$ uncertainty on the $S_{\rm H\alpha}$ from \citet{kaper1997}, and a $10\%$ assumed uncertainty for \citet{gvaramadze2018}, who do not report an uncertainty in their work. Measuring the intersections with the red MeerKAT curve, we find $n_e = 18^{+5}_{-2}$ cm$^{-3}$ and $T = 1.6_{-1.0}^{+4.0}\times10^3$ K using \citet{kaper1997}, and $n_e = 33^{+1}_{-2}$ cm$^{-3}$ and $T = 3.3^{+0.9}_{-0.6}\times10^4$ K using \citet{gvaramadze2018}. 

To assess whether these inferred values are realistic, we can perform several checks. Firstly, we can briefly consider the inferred temperature. Assuming mass, momentum, and energy conversion, we can obtain a post-shock temperature estimate from $kT \sim (3/16)\mu m_p v_*^2$, where $\mu \approx 0.6$ for cosmic abundances and $m_p$ is the proton mass \citep{helder2009}. Using the stellar velocity of Vela X-1, we infer $T \approx 4\times10^4$ K, indeed close to the value inferred using the \citet{gvaramadze2018} H$\alpha$ maps. Secondly, the inferred temperature values are also similar to the range of $6\times10^{3}$--$1.4\times10^{5}$ K found for a sample of H$\alpha$-detected bow shocks in \citet{brown2005}.

Thirdly, we can use the detection of other large scale, diffuse emission structures with MeerKAT. For instance, the ridge of diffuse emission to the South East of Vela X-1, clearly visible in Figure \ref{fig:fullfield}, is also detected in the SuperCOSMOS H$\alpha$ image. In fact, \citet{gvaramadze2018} suggest through simulations that this diffuse structure corresponds to an over-dense region in the ISM, finding that a factor 3 increase in density compared to the surrounding ISM can explain the observed H$\alpha$ morphology of both the field and the bow shock. In both the MeerKAT radio and SuperCOSMOS H$\alpha$ images, the bow shock shows a higher flux than the ridge, although both are of similar order of magnitude. Therefore, one can imagine a scenario where the radio and H$\alpha$ emission of both structures originates from the same (free-free) emission mechanism, with an enhanced bow shock flux caused by a shock-enhanced density increase. 

To test the above scenario, we define a region across the radio-brightest segment of the over-dense ridge in the radio image, as shown by the blue dashed contour in Figure \ref{fig:zoom}. In similar fashion to the bow shock region, we manually defined this arced box region such that it is dominated by ridge emission without containing significant contributions for radio point sources. A slightly different region morphology would, naturally, yield slightly different calculations below; however, this used region samples the mean flux density in the main part of the over-dense ridge. We repeat the bow shock calculation for this over-dense region, measuring a radio flux density of $S_{\nu, \rm overdensity} \approx 28.8$ mJy, integrated over the region's area of $\sim$ 42400 arcsec$^2$. As we don't know the 3D geometry of this over-dense region, we will follow our rationale for the bow shock and assume its depth is similar to its width of $\sim 1.8$ arcmin $\approx 2.1$ parsec \citep[assuming a similar distance as Vela X-1, supported by the evidence for the interaction between Vela X-1 and the over-dense region in][]{gvaramadze2018}.

Using these geometrical assumptions and the flux measurement in Equation \ref{eq:3}, we plot the radio constraints on $n_e$ and $T$ for both the bow shock and the over-dense region in the right panel of Figure \ref{fig:ne_from_both}. The blue and grey shaded areas indicate the temperature ranges derived earlier for the bow shock. Under the assumption that the temperature is not significantly changed in the shock, we find that a shock density increase by a factor $\sim 2.3$ would explain the data. Since, in reality, the temperature in the shock will change as well, the exact density increase will be slightly different -- although we stress that the exact factor scales linearly with the assumed depth of the over-dense region along the line of sight, which has not been measured. 

In adiabatic shocks, the Rankine-Hugoniot equations imply a density enhancement by a factor of 4 \citep{landau1959}. In the analysis by \citet{gvaramadze2018}, a higher density increase, by factors up to $\sim 9$, is inferred from simulations that provide good matches to the observed SuperCOSMOS images. With such enhancement factors, the inferred shock electron densities would imply ISM densities in the range $\sim 2.0$--$4.5$ cm$^{-3}$ \citep[using][]{kaper1997} or $\sim 3.7$--$8.3$ cm$^{-3}$ \citep[using][]{gvaramadze2018}. Literature estimates of the ISM density have been made either using measurements of the standoff distance and stellar (wind) properties, or the optical and IR emission of large scale, close-by HII regions. The former estimates are typically of the order of $\sim 1$--$2$ cm$^{-3}$ \citep[e.g.][]{kaper1997,peri2015,gvaramadze2018}, with higher estimates up to $\sim 10$ cm$^{-3}$ using different stellar wind parameters \citep[e.g.][]{gvaramadze2011}. The latter type tends to suggest higher ISM densities as well, of the order $\sim 5$--$15$ cm$^{-3}$ \citep{kaper1997,lequeux2005}. Therefore, given this range in estimates, the ISM density inferred by combining radio and H$\alpha$ maps appears reasonable, as does the inferred shock density increase (Fig. \ref{fig:ne_from_both}, right). We also note that our inferred densities scale with bow shock depth as $\Delta^{-0.5}$; therefore, a larger depth would imply lower inferred shock and ISM densities. 

Finally, we note that the thermal emissivity scales as $n_e^2$; therefore, for shock density enhancements of at least $4$ to $9$, we expect thermal radio flux densities from the ambient ISM (i.e. outside the over-dense region) that are one to two orders of magnitude fainter (assuming, simplistically, similar temperatures and depths). Such flux densities are below our radio RMS sensitivity, which can explain why the diffuse radio emission so closely traces the over-dense H$\alpha$ region but no other diffuse radio emission is detected.

From the above considerations, the thermal/free-free scenario offers a consistent explanation for the observed emission in the absence of a spectral index measurement. Building on this scenario, we can also make several predictions. Firstly, we can estimate the optical depth in the MeerKAT band, for the typical inferred shock temperatures. The optical depth is given by the integral of the absorption coefficient along the line of sight \citep{longair2011}:

\begin{equation}
    \tau_\nu = \int \chi_\nu dx \approx \chi_\nu \Delta = \frac{\kappa_\nu c^2}{8\pi h \nu^3}\left(e^{h\nu/kT} - 1\right)\Delta \text{ ,}
\end{equation}

\noindent where we have assumed, again, that the shock is homogeneous to obtain analytic estimates. For a typical density $n_e \approx 10$ cm$^{-3}$ and temperature $T \approx 10^4$ K, we find an absorption coefficient in the MeerKAT band of $\chi_\nu \approx 2\times10^{-5}$ pc$^{-1}$. Therefore, given that $\Delta \approx 0.3$ pc, the free-free emission would be optically thin. In fact, for shock parameters of this order of magnitude, we can always expect the emission to be optically thin in the common radio bands. 

The optically thin free-free spectrum has a spectral index of $\alpha \sim 0.1$, up to a break frequency where $h\nu \sim kT$. For the inferred temperatures, we therefore expect this cutoff to lie in the optical bands or at lower frequencies, consistent with the non-detection of X-ray emission. However, directly detecting either this cutoff, or bow shock continuum emission, in optical bands appears unfeasible; at the low-frequency end of the optical band ($\nu \sim 4\times10^{14}$ Hz), the expected average flux is only $\sim 0.08$ $\mu$Jy arcsec$^{-2}$, or $\sim 26.6$ AB mag arcsec$^{-2}$. At lower frequencies, the spectrum is either dominated by dust (IR) or too faint and extended to be detectable with current sub-mm observatories (i.e. ALMA).

\begin{figure}
\includegraphics[width=\columnwidth]{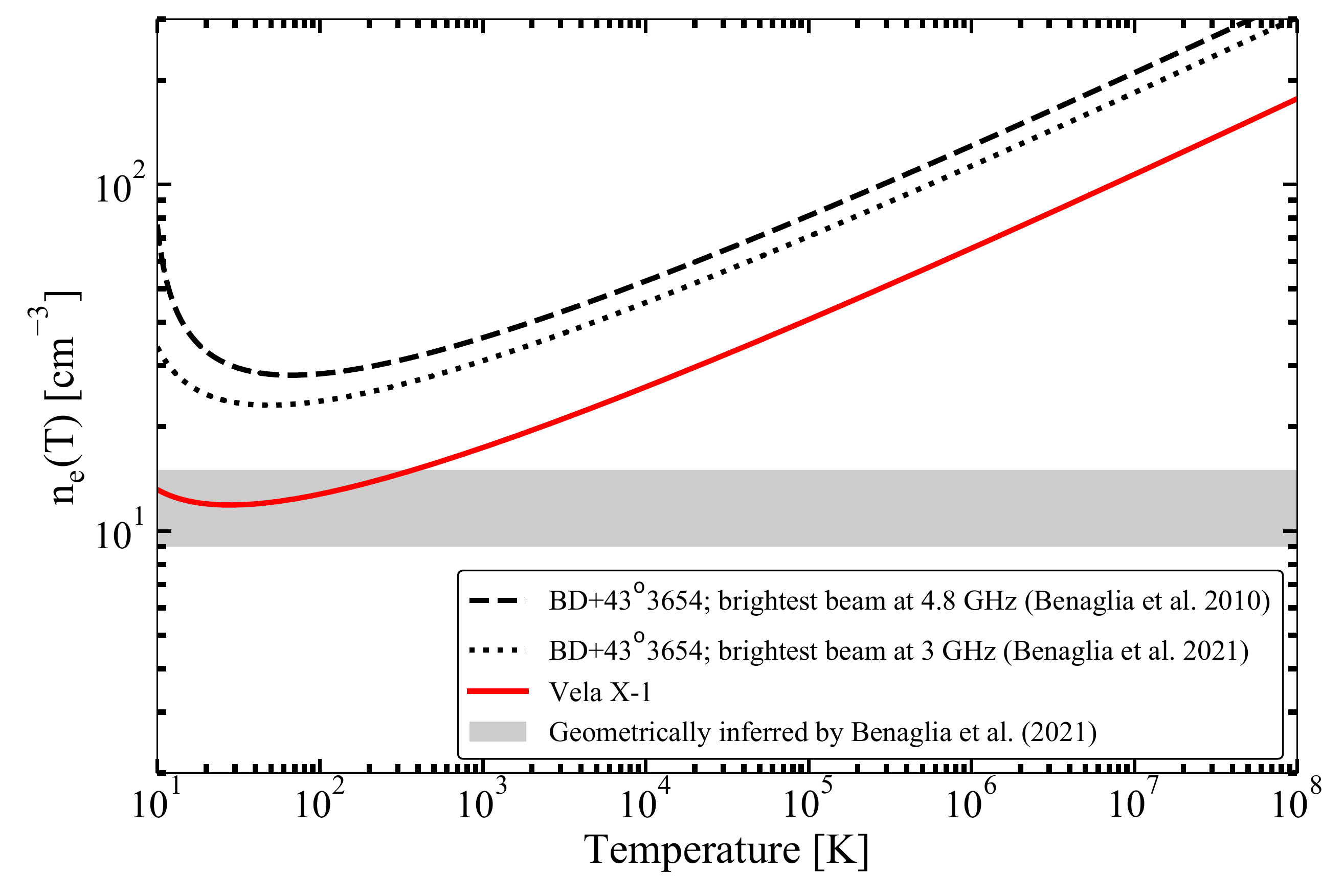}
 \caption{Same as Figure \ref{fig:ne_from_both}, now showing the $n_e(T)$ relation based on the radio detection of Vela X-1 (red) and BD+43$^{\rm o}$3654 (black). For the latter, the two lines indicate estimates based on observations at two different radio frequencies (see text for details). The grey band indicates the inferred ISM electron density ranges discussed in Benaglia et al. (2021).}
 \label{fig:otherbs}
\end{figure}

Finally, we can briefly turn to the only other radio-detected bow shock, BD+43$^{\rm o}$3654, and consider it in the same thermal framework. While its integrated flux density is reported in \citet{benaglia2010} (4.8 GHz) and \citet{benaglia2021}, the total surface area is not. Instead, we can use the flux of brightest beam in the bow shock and consider only that segment of the structure. The radio contours at $4.8$ GHz suggest a peak flux density of the order of $\sim 4$ mJy/bm \citep{benaglia2010}, while at 3 GHz, the peak flux is $\sim 5.5$ mJy/bm \citep{benaglia2021}. At 4.8 GHz, the $12$ arcsec circular beam, $1.7$ kpc distance, and known bow shock width imply a volume of $\sim 4.4\times10^{-2}$ pc$^{3}$. At 3 GHz, the beam size is $20.2$ arcsec x $12.5$ arcsec, implying instead a volume of $\sim 7.7\times10^{-2}$ pc$^{3}$. At both frequencies, we can then combine the flux, volume, and distance estimates to constrain the electron density and temperature. As shown in Figure \ref{fig:otherbs}, both estimates are quite similar, given the simplifying assumptions in this approach. Furthermore, the implied electron densities are between $1.7$--$1.9$ times larger than for Vela X-1. The shock density enhancement, ranges between $3$ to $10$ for temperatures below $\sim 10^5$ K, compared to the ISM density $9$ cm$^{-3}$ around BD+43$^{\rm o}$3654 measured by \citet{benaglia2021}; at a temperature of $\sim 2.6\times10^4$ K, corresponding to a stellar velocity of $43.6$ km/s \citep{benaglia2021} and assuming $kT \sim (3/16)\mu m_p v_*^2$, this factor lies between $6$ and $7$. 

\subsection{The non-thermal/synchrotron scenario}
\label{sec:nonthermal}

Alternatively, the radio emission from the Vela X-1 bow shock could originate from synchrotron emission by a population of relativistic electrons. Such a scenario was first invoked for the radio bow shock in BD+43$^{\rm o}$3654 \citep{benaglia2010} and has since been developed analytically and numerically to predict the radio and X-ray/$\gamma$-ray detectability of other bow shocks \citep[e.g.][]{delvalle2012,delvalle2018,delpalacio2018}. In a nutshell, the stellar wind provides the energy budget to accelerate electrons via diffusive shock acceleration, creating a relativistic electron population. This population loses energy via synchrotron emission, inverse-Compton scattering of dust and stellar photons, and relativistic brehmsstrahlung. Alternatively, particles can escape the bow shock region via diffusion. The relative importance of the processes depends on geometry, magnetic field, and the stellar and IR radiation fields. 

For a subset of the calculations in this Section, we refer the reader to the Appendix for all details. We can start to assess the non-thermal/synchrotron scenario by estimating the shock magnetic field via equipartition arguments, minimizing the combined energy in particles and magnetic fields. In this calculation, and in the remainder of this section, we do not know the radio spectral index $\alpha$. Therefore, we assume that $\alpha \approx 0.5$, as found for BD+43$^{\rm o}$3654 \citep{benaglia2010}. In the Appendix, and several of the figures in this calculation, we will also show the results assuming $\alpha \approx 0.7$. However, we note up front that wherever our qualitative conclusions depend on the exact value of $\alpha$, we will discuss this explicitly.

As detailed in the Appendix, for an equal number of electrons and protons, and assumed minimum and maximum electron energies of $511$ keV and $3\times10^{9}$ keV, respectively, we find an equipartition magnetic field of $B_{\rm eq}~\approx~30$~$\mu$G. The total magnetic and particle energy at equipartition are $W_{\rm mag} \approx 3\times10^{44}$ erg and $W_{\rm par} \approx 4\times10^{44}$ erg. Using the fact that the total wind kinetic power that passes through the bow shock region is given by

\begin{equation}
    L_{\rm wind} = \frac{\eta_{\rm vol} \dot{M}_{\rm wind}  v^2_{\infty}}{2} \approx 5.6\times10^{34} \text{ erg/s,}
    \label{eq:lwind}
\end{equation}

\noindent we find that the time scale to inject the inferred particle energy is roughly $\tau_{\rm eq} \equiv W_{\rm par}/L_{\rm wind} \approx 7\times10^9$ sec. We can then estimate that, without any field amplification, the equipartition magnetic field corresponds to a stellar field $B_{*}$ of the order $B_{*}~\approx~2.5B_{\rm eq} (R_0/R_{*}) \approx 63$ G \citep{benaglia2021,kretschmar2021}. This value is signigicantly lower than the only available Zeeman-splitting measurement for Vela X-1 (known to the authors), which is of the order of several thousand G \citep{bychkov2009}. However, Zeeman splitting was not seen consistently in different lines and was found to be highly time variable \citep{kemp1973,angel1973}. 

Further following the analysis performed for BD+43$^{\rm o}$3654 and other IR bow shocks \citep[][]{delvalle2012,delvalle2018,delvalle2013,delpalacio2018}, we now turn to estimating the relevant time scales for electron acceleration, emission losses, and escape. We consider diffusive shock acceleration as the acceleration process. As radiative processes, we include synchrotron radiation, inverse Compton scattering of electrons in the stellar and IR radiation fields \citep[using the analytic approximations for the Thompson and Klein-Nishina regimes from][]{khangulyan2014}, and relativistic brehmsstrahlung. The full calculations are detailed in the Appendix. 

\begin{figure}
\includegraphics[width=\columnwidth]{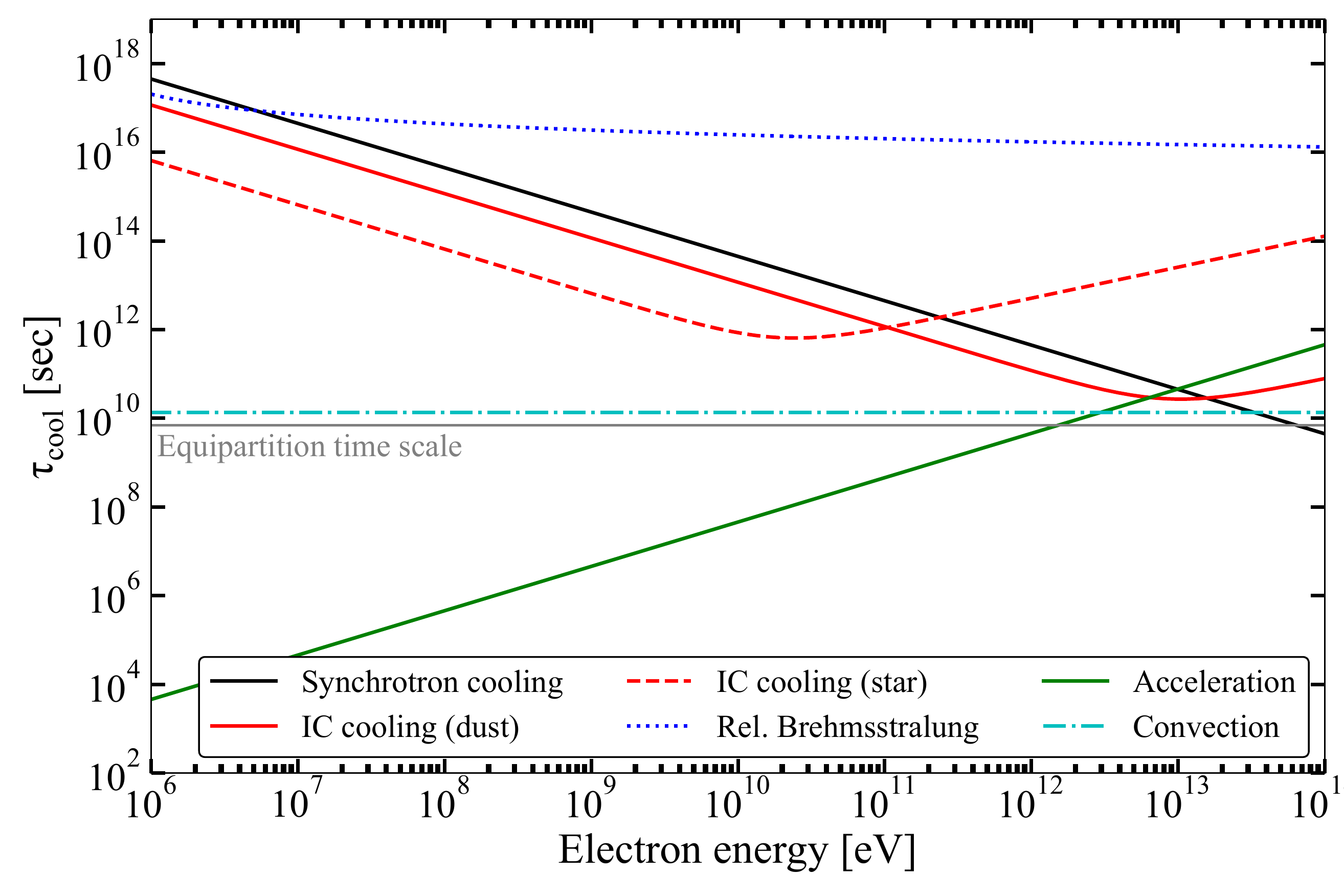}
 \caption{The time scales of electron acceleration, radiative cooling (synchrotron, inverse Compton, relativistic Brehmstrahlung), and convective escape in the bow shock, assuming the equipartition magnetic field $B_{\rm eq} = 30$ $\mu$G. The full calculations are detailed in the Appendix.}
 \label{fig:timescales}
\end{figure}

The resulting time scales are plotted, as a function of electron energy and assuming the equipartition magnetic field, in Figure \ref{fig:timescales}. The maximum electron energy is given by the intersection between the acceleration and shortest loss time scale, the latter being dominated by convective escape. We find a maximum energy around $E_{\rm max} \sim 3\times10^{12}$ eV, similar to our previous assumption. Radiative losses would only become dominant at energies roughly one order of magnitude higher. One notable feature is the similarity of the escape time scale and the equipartition time scale, the former only a factor $\sim 2$ larger than the latter for this magnetic field. In other words, roughly half of all available kinetic wind energy should be injected into particle acceleration to ensure a steady-state scenario -- a required efficiency that further increases when radiative losses become relevant close to $E_{\rm max}$. We will return to this issue of exceptionally high electron acceleration efficiency at the end of this section, fully considering its magnetic field dependence. 

Next, we can consider the broadband synchrotron spectrum to match the observed radio flux density and \textit{Chandra} X-ray flux upper limit. The latter non-detection might observationally constrain the maximum electron energy and magnetic field, as the synchrotron cooling break depends on both those quantities. As derived in the Appendix, $E_{\rm max}$ is set by electron escape for magnetic fields up to a transitional value of $\sim 84$ $\mu$G, above which synchrotron losses take over and determine $E_{\rm max}$ instead. Importantly, this implies that the synchrotron cooling break frequency quadratically increases with magnetic field strength for $B<84$ $\mu$G, while it remains constant at $\nu_{\rm cool} \approx 6\times10^{16}$ Hz for stronger magnetic fields. This maximum cooling break frequency lies a factor two below the minimum energy in the \textit{Chandra} band ($0.5$ keV). Given the steepening of the synchrotron spectrum above the cooling break, this small factor implies that the magnetic field may indeed be constrained by the \textit{Chandra} non-detection. 

\begin{figure}
\includegraphics[width=\columnwidth]{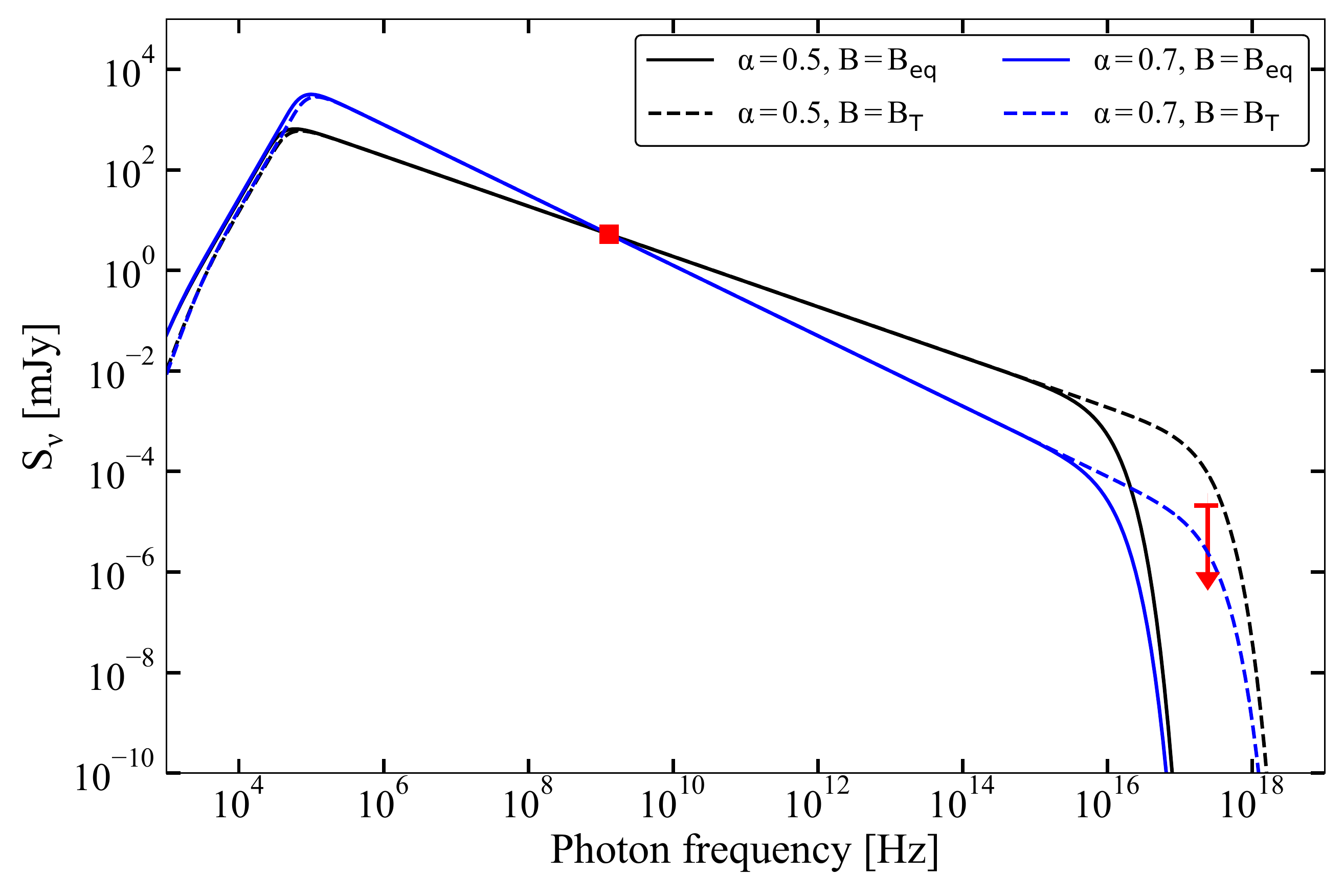}
 \caption{The modelled synchrotron SED for four combination of magnetic field and radio spectral index $\alpha$, normalized to match the observed MeerKAT radio flux. Only for magnetic fields exceeding the maximum field strength implied by the requirement of compressibility of the stellar wind (Equation \ref{eq:compress}) and relatively shallow spectral indices (i.e. the black dashed line), the SED becomes inconsistent with the \textit{Chandra} non-detection.}
 \label{fig:sync_spec}
\end{figure}

In Figure \ref{fig:sync_spec}, we show the expected synchrotron spectra assuming either $\alpha = 0.5$ or $\alpha = 0.7$, and the magnetic field from either equipartition or the transition between the escape and synchrotron dominated regime (both consistently recalculated in the case of $\alpha=0.7$). From these modelled SEDs, we can conclude that the synchrotron spectrum violates the \textit{Chandra} non-detection for magnetic field strengths in the synchrotron-dominated regime, combined with relatively shallow spectral indices (here, $\alpha \sim 0.5$). In other words, the magnetic field should either be closer to equipartition or the spectral index should be steeper. Conversely, for $\alpha \sim 0.7$, any arbitrarily high magnetic field will remain consistent with the \textit{Chandra} non-detection. 

We noted earlier how, under the assumption of equipartition, a significant fraction ($\sim50$\%) of available kinetic wind power would need to be injected into particle acceleration to sustain the system's steady state. This efficiency, however, depends on magnetic field and spectral index: a higher magnetic field implies higher emissitivity of the electrons, requiring a smaller total energy in electrons and a lower acceleration efficiency to match the observed radio flux. Similarly, the spectral index sets the slope of the electron density distribution, and therefore the total energy in electrons. We can define the injection efficiency as 

\begin{equation}
    \eta_e \equiv \frac{\epsilon_e}{L_{\rm wind}} = \frac{V_{\rm bowshock} \int_{E_{\rm min}}^{E_{\rm max}} Q(E) E dE}{L_{\rm wind}} \text{ ,}
    \label{eq:eta1}
\end{equation}

\begin{figure*}
\includegraphics[width=\textwidth]{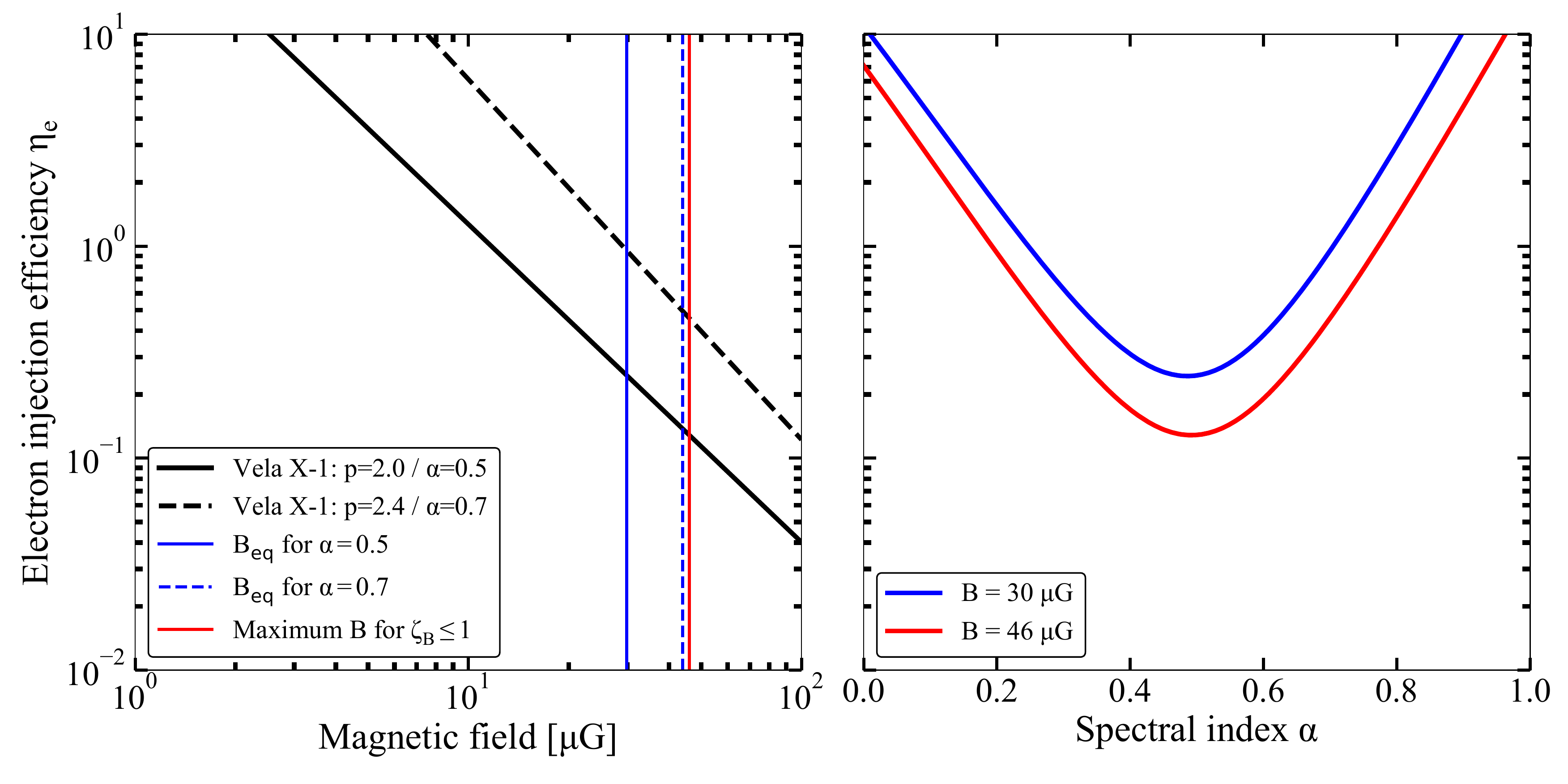}
 \caption{The injection efficiency of electrons, defined in Equation \ref{eq:eta}, as a function of magnetic field (\textit{left}) and radio spectral index (\textit{right}). In the left panel, the efficiency is plotted for different assuming spectral indices, while in the right panel, two different magnetic fields are assumed. The minimum efficiency is obtained for $\alpha = 0.5$ and $B=46$ $\mu$G (i.e. the maximum field strength that leaves the wind compressible), yielding $\eta_e \approx 13$\%.} 
 \label{fig:eff}
\end{figure*}

\noindent where $Q(E)$ is the injection spectrum and $L_{\rm wind}$ is given by Equation \ref{eq:lwind}. As we have seen, the dominant loss mechanism is escape. Since this mechanism is electron-energy-independent, we can approximate the steady state injection spectrum as $Q(E) \approx N(E)/\tau_{\rm esc}$\footnote{See Equation 15 in \citet{delvalle2012} for the full equation, including radiative losses; ignoring those losses slightly underestimates the injection efficiency, but allows for a fully analytic approximation.}, where the electron number density function $N(E)$ is defined in the Appendix. Rewriting the normalisation of $N(E)$ to match the observed radio flux yields (see the Appendix for the complete derivation):

\begin{equation}
    \begin{split}
    \eta_e \approx &\frac{128\pi^3 R_0^3 D^2 S_\nu \epsilon_0 c m_e}{3\sqrt(3)\dot{M}_{\rm wind} v_{\infty} \Delta V_{\rm bowshock} e^3 B a(p)}\\
    &\times \left(\frac{3eB}{2\pi\nu m_e^3 c^4}\right)^{-(p-1)/2} \int_{E_{\rm min}}^{E_{\rm max}} E^{1-p}dE \text{ ,}
    \end{split}
    \label{eq:eta}
\end{equation}

\noindent where $p$ is the power law index of the electron number density distribution, related to the spectral index as $p = 2\alpha + 1$. From this equation, we can conclude that the injection efficiency depends on geometrical properties (e.g. $R_0$, $D$, $V_{\rm bowshock}$), radio observables ($S_\nu$, $p$ in the case of multi-band observations), stellar wind properties, and the magnetic field. We plot the acceleration efficiency as function of magnetic field for $\alpha=0.5$ and $\alpha=0.7$ in Figure \ref{fig:eff} (left). 

Assuming equipartition for the two plotted cases of $\alpha$, we find injection efficiencies of $\eta_e = 25$\% ($\alpha=0.5$) and $\eta_e = 49$\% ($\alpha=0.7$; $B_{\rm eq} = 44$ $\mu$G). Since the system is not necessarily in equipartition, we can alternatively consider higher magnetic fields, corresponding to lower efficiencies. However, the magnetic field cannot be arbitrarily high: firstly, this would greatly increase the energy stored in the magnetic field, and secondly, a shock only forms as long as the magnetic pressure does not exceed the thermal pressure and the wind becomes incompressible. The latter requirement is often parameterized \citep[e.g.][]{delpalacio2018,benaglia2021} through the parameter $\zeta_B$, via 

\begin{equation}
    \frac{B^2}{8\pi} = \zeta_B P = \zeta_B \frac{2}{1+\gamma_{\rm ad}} \rho_{\rm wind} v^2_{\infty} \text{ ,}
    \label{eq:compress}
\end{equation}

\noindent where $\gamma_{\rm ad} = 5/3$ is the ideal gas adiabatic coefficient and $\rho_{\rm wind}$ is the wind density at the stand-off distance. The wind remains compressible, allowing for shock formation, when $\zeta_B \leq 1$, implying a maximum magnetic field of $B\approx 46$ $\mu$G. At this magnetic field, shown as the red line in Figure \ref{fig:eff} (left), the implied injection efficiencies are 13\% and 45\% for $\alpha=0.5$ and $\alpha=0.7$, respectively. 

These values for the injection efficiency are significantly higher than inferred in comparable systems: for instance, \citet{delpalacio2018} infer values between $16$\% and $0.4\%$ assuming $\zeta_B = 0.01$ and $\zeta_B = 1$, respectively, for BD+43$^{\rm o}$3654. Similarly applying two extreme values of $\zeta_B$, \citet{benaglia2021} find efficiencies of $10$\% ($\zeta_B = 0.03$) and $1.5$\% ($\zeta_B = 1.0$) for the same source. Indeed, we can apply Equation \ref{eq:eta} to the brightest beam of BD+43$^{\rm o}$3654 in \citet{benaglia2010} (see the end of Section \ref{sec:thermal}) assuming the stellar wind parameters reported in that work: this confirms that the required injection efficiencies in BD+43$^{\rm o}$3654 are one order of magnitude lower for the same spectral index $\alpha$. Further comparisons can be made with \citet{stappers2003}, who report efficiencies of $\sim 4$\% and $<9$\% for the Crab nebula and the pulsar wind nebula around B1957+20, respectively.

These high injection efficiencies required in the Vela X-1 bow shock are not the result of assuming incorrect values of $\alpha$ either. While $\alpha$ might be different than $0.5$ or $0.7$, the injection efficiency does not depend monotonically on spectral index. In the right panel of Figure \ref{fig:eff}, we plot $\eta_e$ versus $\alpha$ for two different magnetic field (equipartition for $\alpha = 0.5$ and assuming $\zeta_B = 1$). In both cases, the efficiency is minimized around $\alpha = 0.5$. One can also see how a higher magnetic field lowers the efficiency, implying that $\eta_e(B=B_{\rm max}, \alpha=0.5) = 13$\% is the lowest possible value for Vela X-1. Finally, for significantly lower $\alpha$, the non-detection of X-ray emission requires a lower magnetic field (e.g. Figures \ref{fig:break} and \ref{fig:sync_spec}), further increasing the inferred injection efficiency.

We can conclude that only a fine-tuned combination of the electron density distribution slope and magnetic field yields an injection efficiency approaching the maximum values inferred for BD+43$^{\rm o}$3654 and other types of nebula. Instead, without such fine-tuning, significantly higher values up to 100\% are required to explain the Vela X-1 bow shock radio emission fully through a non-thermal scenario. 

Alternatively, we can turn this argument around: if the injection efficiency is in reality a factor ten lower than those derived above (i.e. $\sim 1.3$\% at minimum), one tenth of the observed radio flux could be attributed to synchrotron processes. As the electron densities in the thermal scenario scale with $S_\nu^{0.5}$, this would imply only a $\sim 5$\% decrease in the inferred $n_e$. Such a decrease does not invalidate the arguments in Section \ref{sec:thermal}, and is smaller than the systematic differences between the densities inferred using the \citet{kaper1997} and \citet{gvaramadze2018} H$\alpha$ maps. Therefore, a dual thermal + non-thermal scenario, with the observed flux dominated by the former, is a realistic possibility.

\subsection{Why are only the bow shocks of BD+43$^{\rm o}$3654 and Vela X-1 detected in radio?}

The MeerKAT detection of the Vela X-1 bow shock is only the second radio bow shock of a runaway massive star after the discovery of the prototype in BD+43$^{\rm o}$3654 \citep{benaglia2010}, and the first radio bow shock around an HMXB. Radio searches for other bow shocks have since been performed using the NRAO VLA Sky Survey \citep[NVSS;][]{condon1998} for HIP 16518, HIP 34536, HIP 78401, HIP 97796 \citep{debecker2017} and pointed VLA observations \citep[as mentioned by][based on private communication with C. Peri]{benaglia2021}. However, these searches have all remained unsuccessful. Considering both the thermal or non-thermal scenario for the radio emission (or their combination), we will now briefly discuss the future prospects for radio bow shock detections and whether exceptional physical conditions are required.

Firstly, we can consider the non-thermal scenario. The four stars that \citet{debecker2017} did not detect in radio, can be compared to Vela X-1 and BD+43$^{\rm o}$3654 in the frame work introduced in Section \ref{sec:nonthermal}. For these sources, \citet{debecker2017} collect distances, bow shock widths, stellar mass loss rates, and wind velocities. For these, we can assume a maximum electron energy of $10^{12}$ eV and a high injection efficiency of $\eta_e = 10$\%. Using these numbers, we can estimate the expected flux density in a $1.4$-GHz, $45$ arcsec beam of the NVSS. For these estimates, we assume a best-case scenario of $\alpha=0.5$, corresponding to the highest $S_\nu$ for a given $\eta_e$ (Figure \ref{fig:eff}, right), and consider two high values of $B$ ($50$ and $100$ $\mu$G). In all cases, the bow shock is larger than the beam size.

For all four stars, non-thermal flux densities in the range of 0.4--1.5 mJy/beam and 1.0--4.3 mJy/beam can be expected, assuming $B = 50$ and $100$ $\mu$G, respectively. Given the typical RMS of the survey of $\sim 0.5$ mJy/beam, the non-detection of radio emission from these bow shocks is unsurprising: only in the best-case circumstances in terms of spectral shape, magnetic field, and injection efficiency, only HIP 34536 might be three-sigma detectable for the strongest considered magnetic field. However, even then, at such low significance, likely no bow shock morphology will be identifiable. These sources are not outliers within the E-BOSS bow shock samples, nor are they similar in their relevant properties: they span wind velocities between $500$ and $2500$ km/s, mass loss rates between $6\times10^{-9}$ to $5\times10^{-7}$ $\rm M_{\odot}$/yr, and distances between 0.22 and 2.2 kpc. Therefore, we deem it reasonable to extend this conclusion to the wider bow shock sample. 

Similarly, we can investigate a thermal scenario. For all four systems considered above, the first E-BOSS catalogue lists inferred ISM densities. Using the same geometrical assumptions as in the non-thermal scenario, and assuming a temperature, we can again predict the expected radio flux in the NVSS. For these four systems, the ISM densities range between $0.01$ and $2$ cm$^{-3}$. We can assume a shock density enhancement of $4$ and, more extremely, $10$ (Gvaramadze et al. 2018). However, we estimate flux densities of only $0.2$ $\mu$Jy/beam to $0.8$ mJy/beam in the latter case, assuming a temperature of $T = 10^4$ K. Even at a lower temperature of $10^3$ K, the maximum flux density is $1.85$ mJy/beam, for HIP 78401, which has the highest inferred $n_e$ and smallest distance. Therefore, we also do not expect any significantly detectable radio emission in the NVSS in the thermal scenario.

In the thermal scenario, the most effective band to search for emission depends on the electron temperature. Comparing Equations \ref{eq:freefree} and \ref{eq:halpha}, it is clear that the ratio of radio flux density and H$\alpha$ surface brightness is independent of the electron density. However, their ratio will depend strongly on temperature, i.e. $S_{\nu} / S_{\rm H\alpha} \propto T^{0.4}g(\nu, T)$ (ignoring the exponent in Equation \ref{eq:freefree}). In the temperature range between $10^2$ and $10^5$ K, this ratio increases by a factor $50$, implying that the radio continuum emission may be easier to pick up at higher temperatures. However, the absolute flux density/surface brightness depends very strongly on electron density as well; a relatively hot but low density medium will remain undetectable in both bands.

What then makes BD+43$^{\rm o}$3654 and Vela X-1 stand out? For Vela X-1, the high sensitivity and core-heavy configurations of MeerKAT are vital to detect the bow shock's radio emission. However, as discussed in Section \ref{sec:thermal}, the presence of a local overdense ISM structure \citep{gvaramadze2018} is as essential in a free-free scenario. In BD+43$^{\rm o}$3654, on the other hand, both a relatively high local ISM density and a high kinetic wind power could be the crucial factor: in our injection efficiency estimates, we assumed a stellar mass-loss rate of $1.6\times10^{-4}$ $\rm M_{\odot}$/yr and a velocity of $2300$ km/s \citep{kobulnicky2010,benaglia2010}, leading to significantly higher wind powers than in the four stars considered above. It may be that, indeed, either exceptional stellar wind or ISM properties are essential for a radio bow shock detection. 

The prospect of recent and upcoming all-sky radio surveys has several consequences for radio bow shock studies. One example of such a survey is the aforementioned RACS \citep{mcconnell2020} that covers the entire sky South of $+40^{\rm o}$ declination at $887.5$ MHz down to a typical RMS sensitivity of $0.25$ mJy/bm. For RACS, the $\sim 10$ times smaller beam area combined with the $\sim 2$ times better sensitivity implies that the flux detectability of individual bow shocks (as considered above) does not greatly improve compared to NVSS. However, given the uniform sky coverage, especially in the Galactic plane and centre regions, a large number of known bow shocks can be investigated further in the radio band. The ongoing Galactic plane survey with MeerKAT ({\color{blue} Goedhart et al., in prep.}) will provide similar sky coverage but at higher sensitivity and, as our detection of the Vela X-1 bow shock shows, with an array configuration especially suitable for bow shock studies. Crucially, with both surveys, any bow shock that is radio detected will be better resolved than it would have been with the NVSS, helping to distinguish between bow shocks and other morphologies. Especially in a thermal scenario with a high surrounding ISM density, comparing the radio morphology to IR maps may be vital to identify radio counterparts. However, based on our considerations, detecting radio bow shocks in these large-scale surveys may still requires either high stellar wind kinetic powers (such as BD+43$^{\rm o}$3654) or locally over-dense or complex ISM environments (such as Vela X-1).

\section{Summary and conclusions}

In this paper, we have presented the discovery of $1.3$-GHz radio emission from the Vela X-1 bow shock with MeerKAT. The MeerKAT data also reveal other large scale structures of diffuse radio emission, tracing the known H$\alpha$ structures in the field. Our analysis of publicly available X-ray and lower-frequency radio observations does not reveal bow shock emission at these other frequencies. These results present only the second radio bow shock around a massive runaway star, compared to over 700 known bow shocks observed in the IR band. It also present the first radio bow shock detected around an X-ray binary. 

Lacking a radio spectral index measurement, we turn to the bow shock's energetics and brightness to assess the underlying emission mechanism. We firstly consider a thermal scenario, dominated by optically thin free-free emission. In such a scenario, the radio and H$\alpha$ emission of both the bow shock and diffuse structures, originates from the same process and particle population. Combining their constraints, we find reasonable estimates for the bow shock and ISM density and temperature, consistent within their systematics with earlier, independent estimates. Alternatively, a non-thermal scenario, dominated by synchrotron emission form a shock-accelerated population of electrons, is harder to reconcile with the observed bow shock properties. In particular, it requires high energy injection efficiencies of $\gtrsim 13$\%, depending on the exact electron distribution and magnetic field. The observed emission may, of course, in reality originate from a combination of both processes, dominated by the thermal emission. 

Finally, we consider why the great majority of stellar bow shocks have escaped detection at radio frequencies so far. Our considerations regarding the thermal and non-thermal scenarios, suggest that either high density/complex ISM environments or exceptionally energetic stellar winds, respectively, are required for such a detection. However, as our MeerKAT observations show, the advent of SKA precursors and the future SKA may allow for the detection of a significant number of other stellar bow shocks. Building up such a sample will test the hypothesis above, and may reveal new examples of synchrotron-dominated bow shocks as evidence of electron acceleration to very high energies.


\section*{Acknowledgements}


We thank the anonymous referee for their helpful report that improved the quality and clarity of this work. We thank the staff at the South African Radio Astronomy Observatory (SARAO) for scheduling these observations. The MeerKAT telescope is operated by the South African Radio Astronomy Observatory, which is a facility of the National Research Foundation, an agency of the Department of Science and Innovation. This work was carried out in part using facilities and data processing pipelines developed at the Inter-University Institute for Data Intensive Astronomy (IDIA). IDIA is a partnership of the Universities of Cape Town, of the Western Cape and of Pretoria. We acknowledge the use of data obtained from the High Energy Astrophysics Science Archive Research Center (HEASARC), provided by NASA’s Goddard Space Flight Center. This paper includes archived data obtained through the CSIRO ASKAP Science Data Archive, CASDA (\url{https://data.csiro.au}). The Australian SKA Pathfinder is part of the Australia Telescope National Facility (grid.421683.a) which is managed by CSIRO. Operation of ASKAP is funded by the Australian Government with support from the National Collaborative Research Infrastructure Strategy. ASKAP uses the resources of the Pawsey Supercomputing Centre. Establishment of ASKAP, the Murchison Radio-astronomy Observatory and the Pawsey Supercomputing Centre are initiatives of the Australian Government, with support from the Government of Western Australia and the Science and Industry Endowment Fund. We acknowledge the Wajarri Yamatji people as the traditional owners of the Observatory site. This work makes use of several \textsc{python} packages, namely \textsc{numpy} \citep{oliphant_numpy}, \textsc{astropy} \citep{astropy13,astropy18}, \textsc{matplotlib} \citep{hunter07}, and \textsc{aplpy} \citep{robitaille12}. JvdE is supported by a Lee Hysan Junior Research Fellowship awarded by St. Hilda's College, Oxford. TDR acknowledges financial contribution from ASI-INAF n.2017-14-H.0, an INAF main stream grant. GRS is supported by NSERC Discovery Grants RGPIN-2016-06569 and RGPIN-2021-04001. SEM acknowledges financial support from the Violette and Samuel Glasstone Research Fellowship programme, the UK Science and Technology Facilities Council (STFC), and Oxford Centre for Astrophysical Surveys, which is funded through generous support from the Hintze Family Charitable Foundation. SEM also acknowledges financial contribution from the agreement ASI-INAF n.2017-14-H.0 and the INAF mainstream grant, and from PRIN-INAF 2019 n.15. S.M. acknowledges funding from the South African National Research Foundation (NRF) and University of Cape Town VC2030 Future Leaders Award. JCAM-J is the recipient of Australian Research Council Future Fellowship (project number FT140101082).

\section*{Data Availability Statement}

A Jupyter notebook to generate the figures in this work and redo the analysis in the Discussion, can be found on this link upon publication: \url{https://github.com/jvandeneijnden/VelaX-1-MKAT}. There, the underlying MeerKAT and \textit{Chandra} images are also included. The raw MeerKAT and \textit{Chandra} data, as well as the \textit{RACS} images, are publicly available in their respective online observatory repositories.

\input{output.bbl}

\appendix

\section{Non-thermal calculations}
\label{sec:appendix}

In this appendix, we will provide addition derivations of the calculations presented in Section \ref{sec:nonthermal} of the main paper. As noted there, the below calculations are build upon the pioneering work of \citet{benaglia2010}, \citet{delvalle2012,delvalle2018}, and \citet{delpalacio2018}. We refer the reader to those works for more details underlying the calculations. 

\subsection{Equipartition and particle energies}

Firstly, let us consider the total energy in particles given an observed radio flux, and an assumed magnetic field and radio spectral index $\alpha$. Following \citet{longair2011}, we start with the common assumption that the electron number density distribution takes a power law form as a function of electron energy $E$: $N(E) = \kappa E^{-p}$ between a minimum and maximum energy $E_{\rm min}$ and $E_{\rm max}$, where $p=2\alpha+1$ and $S_\nu \propto \nu^{-\alpha}$. We set the minimum energy to the electron rest mass, i.e. $E_{\rm min} = 511$ keV. The maximum energy is discussed in the main text and below; for this initial calculation, we will assume a typical value of $E_{\rm max} = 3\times10^{9}$ keV. We then define the ratio between protons and electrons as $a$, such that the total energy in particles is $\epsilon_{\rm particles} = (1+a)\epsilon_{\rm electrons} \equiv \eta \epsilon_{\rm electrons}$. The total energy in electrons is simply the weighted mean of the electron number density multiplied by the bow shock volume $V_{\rm bowshock}$:

\begin{equation}
    W_{\rm electrons} = V_{\rm bowshock} \times \epsilon_{\rm electrons} = V_{\rm bowshock} \int_{E_{\rm min}}^{E_{\rm max}}\kappa E^{1-p} dE \text{ .}
\end{equation}

\noindent The normalisation $\kappa$ is constrained by the radio flux, for a given magnetic field and spectral shape: 

\begin{equation}
    J_{\nu} = \frac{\sqrt{3}e^3 B \kappa}{4\pi\epsilon_0 c m_e} \left(\frac{3eB}{2\pi\nu m_e^3 c^4}\right)^{(p-1)/2} a(p) \text{ ,}
    \label{eq:jnu}
\end{equation}

\noindent where the factor $a(p)$ is defined as

\begin{equation}
    a(p) = \frac{\sqrt{\pi}}{2} \frac{\Gamma\left(\frac{p}{4} + \frac{19}{12} \right) \Gamma\left(\frac{p}{4} - \frac{1}{12} \right) \Gamma\left(\frac{p}{4} + \frac{5}{4} \right)}{(p+1) \Gamma\left(\frac{p}{4} + \frac{7}{4} \right)} \text{ ,}
\end{equation}

\noindent and $\Gamma(x)$ is the Gamma-function. As a final ingredient, the observed radio flux is related to $J_{\nu}$ via $J_{\nu} = 4\pi D^2 S_{\nu} / V_{\rm bowshock}$. Combining these equations into an expression for the total particle energy yields:

\begin{equation}
    W_{\rm par} = 4\pi D^2 S_{\nu} \eta \frac{4\pi\epsilon_0 c m_e}{\sqrt{3}e^3 B a(p)} \left(\frac{3eB}{2\pi\nu m_e^3 c^4}\right)^{-\frac{p-1}{2}} \int_{E_{\rm min}}^{E_{\rm max}} E^{1-p} dE \text{ .}
\end{equation}

\noindent The energy in the magnetic field is, on the other hand, given by 

\begin{equation}
    W_{\rm mag} = V_{\rm bowshock}\frac{B^2}{8\pi} \text{ .}
    \label{eq:wmag}
\end{equation}

\begin{figure}
\includegraphics[width=\columnwidth]{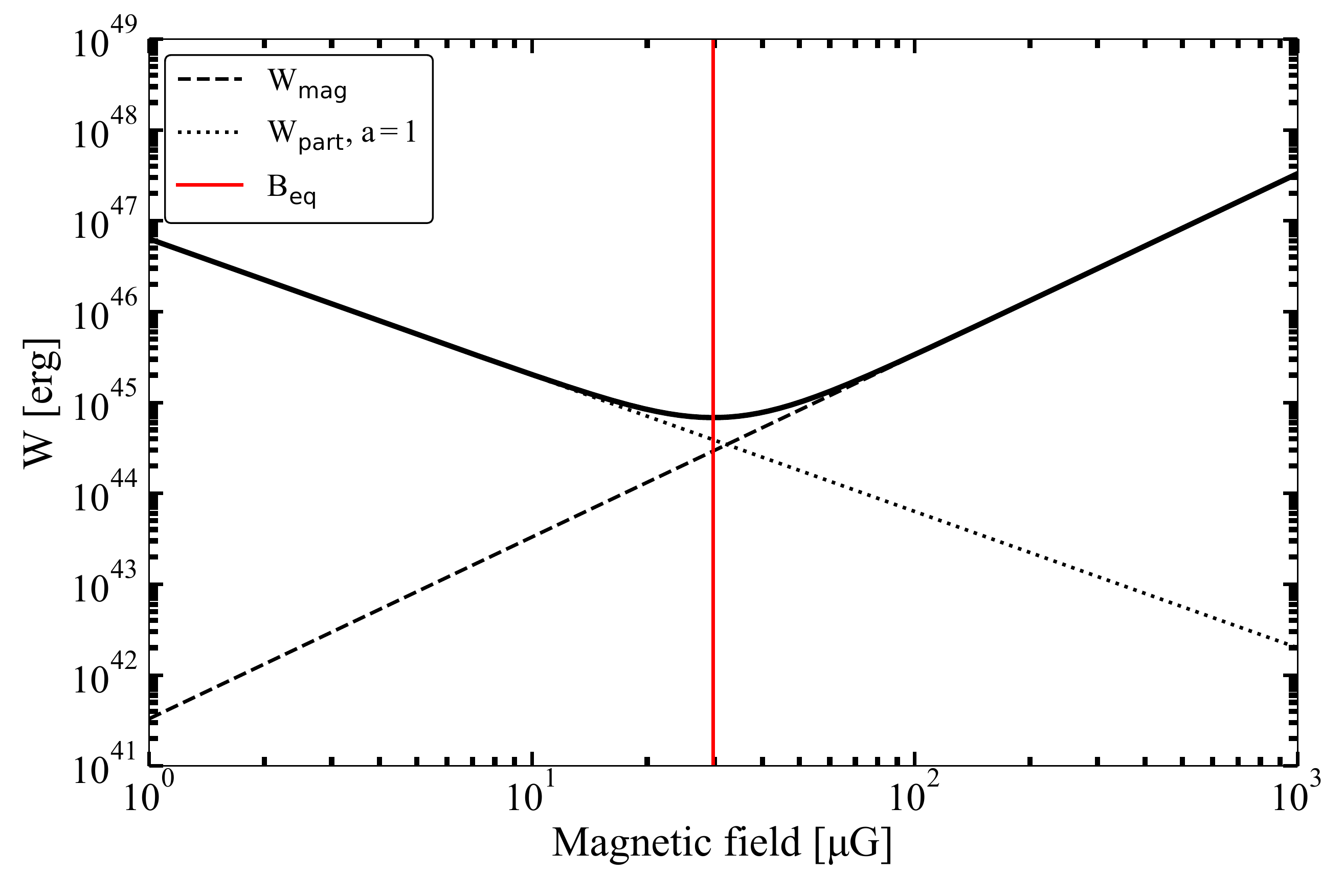}
 \caption{The energy contained in relativistic particles and the magnetic field, as a function of the magnetic field strength, assuming an equal number of protons and electrons. For a spectral index $\alpha = 0.5$, we measure an equipartition magnetic field of $B_{\rm eq} \approx 30$ $\mu$G.}
 \label{fig:eq}
\end{figure}

\noindent Minimizing the sum of the two energies as a function of magnetic field, as plotted in Figure \ref{fig:eq}, we find an equipartition magnetic field of $B_{\rm eq}~\approx~30$~$\mu$G. The total magnetic and particle energy at equipartition are $W_{\rm mag} \approx 3\times10^{44}$ erg and $W_{\rm par} \approx 4\times10^{44}$ erg. 

\subsection{Loss and acceleration time scales}

We calculate the cooling time scales following earlier calculations by \citet{delvalle2012,delvalle2018} and \citet{delpalacio2018}. Firstly, the inverse synchtrotron cooling time scale is given by

\begin{equation}
    t_{\rm sync}^{-1} = \frac{4}{3}\frac{\sigma_{\rm T} c W_{\rm mag}}{m_e c^2} \left(\frac{E}{m_e c^2}\right) \text{ ,}
\end{equation}

\noindent where $\sigma_{\rm T}$ is the Thompson cross-section and the magnetic field energy density is given by Equation \ref{eq:wmag}. Then, turning to the inverse-Compton losses, we can consider upscattering of either IR or stellar photons. We will follow the approximations in \citet{khangulyan2014} to calculate the time scales, assuming a blackbody photon spectrum, in both the Thompson (low electron energy) and Klein-Nishina (high electron energy) limits. The transition between these two limits is set by the characteristic energy of the ambient photon field (more precisely, when the electron Lorentz factor reaches above $\gamma \leq h\nu_0 / (4m_e c^2)$ where $h\nu_0$ is the characteristic photon energy). For the inverse-Compton scattering with dust photons, we first estimate the dust temperature following \citet{delvalle2012} and \citet{draine1981}, as

\begin{equation}
    T_{\rm dust} \approx 27 \left(\frac{a_{\rm dust}}{\mu\text{m}}\right)^{-1/6} \left(\frac{L_{*}}{10^{38} \text{ erg/s}}\right)^{1/6} \left(\frac{R_0}{\text{pc}}\right)^{-1/3} \text{ K} \approx 73\text{ K,} 
\end{equation}

\noindent where the typical dust radius is assumed to be $a_{\rm dust} \approx 0.2$ $\mu$m and the stellar luminosity $L_{*} = 24.3\times10^{38}$ erg/s \citep{kretschmar2021}. We also calculate a grey-body (or dilution) correction $\kappa_{\rm dust}$ using the framework introduced by \citet{debecker2017} and \citet{delpalacio2018}, based on the observed, integrated IR magntiude (5.25 mag; $L_{\rm IR} \approx 9.1\times10^{36}$ erg/s) and 

\begin{equation}
    \kappa_{\rm dust} = \frac{L_{\rm IR}}{A \sigma_{\rm SB} T_{\rm dust}^4} \approx 6.5\times10^{-4} \text{ .}
\end{equation}

\noindent The introduction of this factor accounts for the issue that the dust is not completely optically thick, which causes the significant discrepancy between the observed IR flux and that predicted from $T_{\rm dust}$. With this grey-body correction in hand, we can calculate the cooling time scales using Equations 41 and 42 in \citet{khangulyan2014} for the Thompson and Klein-Nishina regimes, respectively. In these equations, both the electron energy and dust temperature are scaled to the electron rest mass and are denoted by $\overline{E}$ and $\overline{T}_{\rm dust} \equiv kT_{\rm dust}/m_e c^2$:

\begin{equation}
    \left(t_{\rm IC,dust}^{\rm Thompson}\right)^{-1} = \frac{4c\overline{E}}{9}\frac{\pi^2 \kappa_{\rm dust} \overline{T}_{\rm dust}^4 m_e^3 c^3 8\pi r_0^2}{15 \hbar^3}
    \label{eq:ic_T}
\end{equation}

\begin{equation}
    t_{\rm IC,dust}^{\rm KN} = \left(5\times10^{-17}\text{ sec}\right)\times \overline{T}_{\rm dust}^{-2.3} \overline{E}^{0.7} \kappa_{\rm dust}^{-1} \text{ .}
    \label{eq:ic_KN}
\end{equation}

\noindent Finally, the total inverse-Compton time scale is calculated as the inverse of Equation \ref{eq:ic_T} summed with Equation \ref{eq:ic_KN}. We then repeat this calculation for the stellar radiation field, where the dust temperature is replaced by the stellar temperature $T_* = 30.9\times10^3$ K. The dilution factor is now given by Equation 31 in \citet{khangulyan2014}, $\kappa_* = (R_*/2R_0)^2 = 3.5\times10^{-13}$. Due to this low dilution factor, compared to the dust contribution, the effective stellar photon temperature is lower than the dust temperature, resulting in a transition between the two regimes at a lower electron energy. 

Finally, we calculate the relativistic Brehmsstrahlung via Equation 11 in \citet{delvalle2012}, assuming that the wind density is enhanced by a factor $4$ in the shock. The time scale of escape, the final loss mechanism, is energy independent and is estimated as $t_{\rm escape} \approx \Delta / v_{\infty}$. Following common assumptions, we model the acceleration time scale assuming diffusive shock acceleration, given in \citet{terada2012} as (where we correct the factor $c$ to yield the correct units) $t_{\rm acc} = (20\xi/3)\times(E/eB)\times(1/v_{\infty})^2$. We also assume Bohm diffusion, i.e. $\xi=1$. 

\subsection{The broadband synchrotron spectrum}

To model the broadband synchrotron spectrum, we (i) assume a spectral index, (ii) match the observed MeerKAT L-band radio flux density to the SED, and (iii) calculate the low- and high-frequency behaviour due to synchrotron-self-absorption and the maximum electron energy, respectively. We can then also compare this to the X-ray upper limit: this non-detection may specifically constrain the maximum electron energy and magnetic field, as the high-frequency break depends on both those quantities. However, $E_{\rm max}$ and $B$ are themselves also related: at equipartition, Figure \ref{fig:timescales} shows that electron escape sets the maximum energy. For different magnetic field strengths, the acceleration time scale will decrease as $B^{-1}$, while the synchrotron cooling time scale decreases as $B^{-2}$. As escape and inverse Compton processes are all independent of the magnetic field, at some $B$, synchrotron losses will set the maximum energy and, in turn, the cooling break. 

Where diffusive losses are dominant, we can find the maximum electron energy by equating the escape and acceleration time scales:

\begin{equation}
    E_{\rm max} = \frac{3\Delta v_{\infty} e}{20\xi} B \equiv C_{\rm esc} B \text{ .}
\end{equation}

\noindent If, instead, at a high magnetic field strength, the synchrotron radiative losses dominate, we can instead derive (again equating the time scales):

\begin{equation}
    E_{\rm max} = 3 m_e c^2 v_{\infty} \sqrt{\frac{\pi e}{10\xi\sigma_{\rm T} c}} B^{-1/2} \equiv C_{\rm sync}B^{-1/2} \text{ .}
\end{equation}

\noindent Comparing both equations, the transition between the escape and synchrotron dominated cases occurs at a magnetic field of $B_{\rm T} = (C_{\rm sync}/C_{\rm esc})^{2/3}$. For Vela X-1, $B_T \approx 84$ $\mu$G, larger than the maximum magnetic field to allow the wind to be compressible. Therefore, the Vela X-1 bow shock is expected to be escape dominated. 

The cooling break in the synchrotron spectrum can then be approximated by the characteristic emission frequency for electrons at the maximum energy, i.e.

\begin{equation}
\nu_{\rm cool} = \frac{eB}{2\pi m_e}\left(\frac{E^{\rm max}_{\rm electron}}{m_e c^2}\right)^2 \text{ ,}
\end{equation}

\noindent where, clearly, the magnetic field dependence disappears in the synchrotron-dominated case ($E_{\rm max} \propto B^{-0.5}$). Introducing the maximum energy for both regimes into the above equation yields:

\begin{equation}
\begin{split}
B \leq B_T \text{: } \nu_{\rm cool} &= \frac{e C_{\rm escape}^2 B^3}{2\pi m_e}\left(\frac{1}{m_e c^2}\right)^2 \\
B > B_T \text{: }\nu_{\rm cool} &= \frac{e C_{\rm sync}^2}{2\pi m_e}\left(\frac{1}{m_e c^2}\right)^2 \text{ .}
\end{split}
\end{equation}

\begin{figure}
\includegraphics[width=\columnwidth]{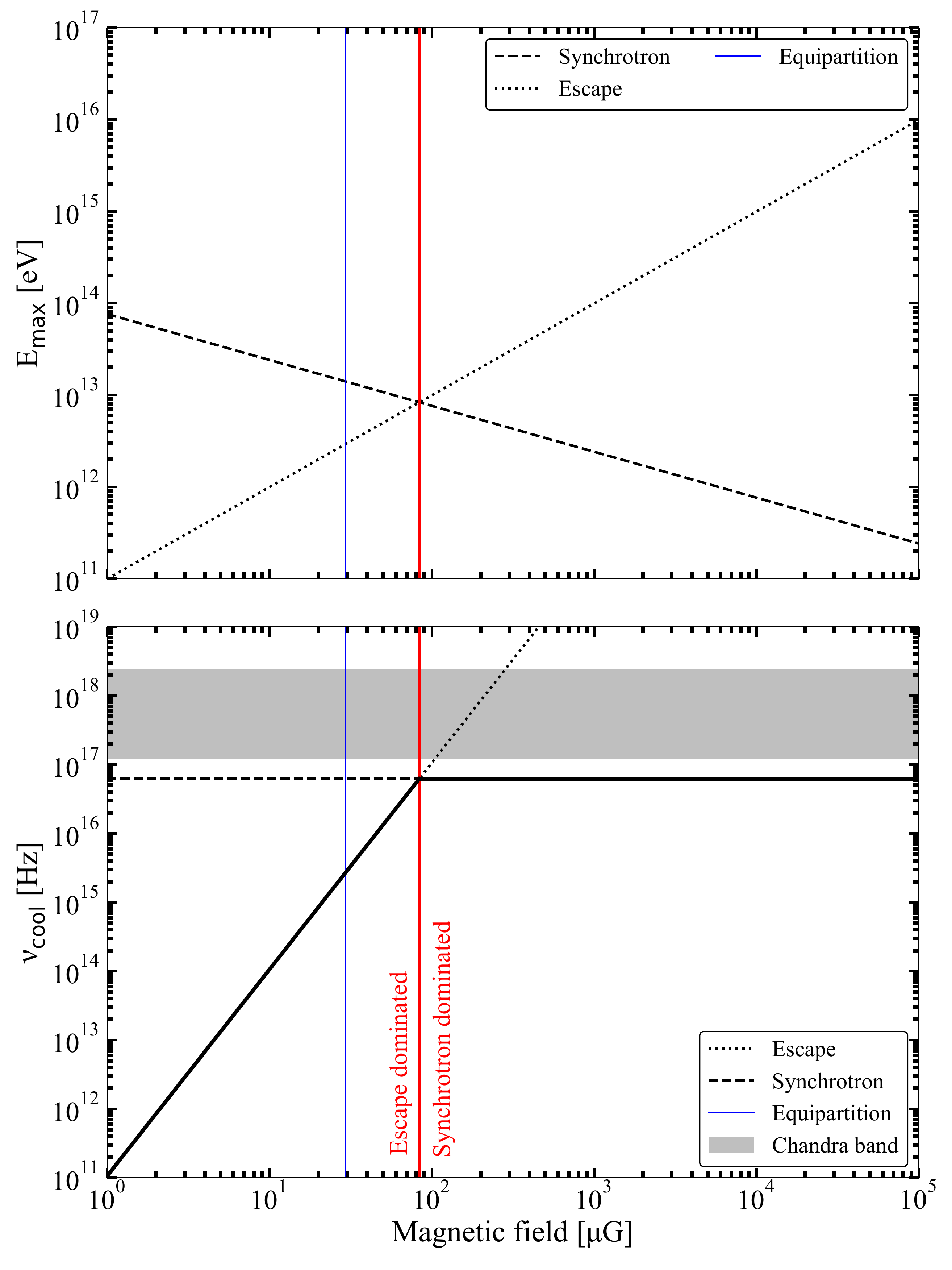}
 \caption{The maximum electron energy (\textit{top}) and resulting synchrotron cooling break frequency (\textit{bottom}) as a function of magnetic field strength. \textit{Escape} and \textit{synchrotron} indicate the dominant energy loss mechanism, setting the maximum electron energy. At no point does the cooling break frequency extend into or above the \textit{Chandra} band.}
 \label{fig:break}
\end{figure}

\noindent We plot these relations in the top panel of Figure \ref{fig:break}. Because the magnetic field dependence drops out for the synchrotron-dominated regime, we find a maximum cooling break frequency of $\nu_{\rm cool} \approx 6\times10^{16}$ Hz, regardless of spectral index. Having derived the cooling break frequency, we can model the overall photon-energy-dependent shape of the spectrum by following \citet{delvalle2012}\footnote{Note that this equation corrects a typo in the original reference, adding the missing minus in the exponent.}, e.g.

\begin{equation}
    F_{\rm sync}(E_{\gamma}) \propto \kappa_{\nu, \rm SSA} \int_{E_{\rm min}}^{E_{\rm max}} N(E) \left(\frac{E_{\gamma}}{E_c(E)}\right)^{4/3}e^{-\frac{E_{\gamma}}{E_c(E)}}dE \text{ ,}
\end{equation}

\noindent where the synchrotron-self-absorption coefficient is defined as $\kappa_{\nu, \rm SSA} = (1-e^{-\tau_\nu})/\tau_\nu$ and $\tau_\nu$ is given by Eq. 8.132 in \citet{longair2011}. Moreover, $E_\gamma$ is the photon energy, and the critical energy is defined by \citet{delvalle2012} as

\begin{equation}
    E_c(E) = \frac{3}{4\pi}\frac{ehB}{mc}\left(\frac{E}{mc^2}\right)^2 \text{ .}
\end{equation}

\noindent Finally, as stated, the spectrum is normalised such that it matches the observed integrated radio flux density. Using this approach, we plot the four examples in Figure \ref{fig:sync_spec}.

\subsection{The injection efficiency}

As defined in the main text, the injection efficiency is calculated as the energy injected into electrons divided by the available kinetic power from the stellar wind:

\begin{equation}
    \eta_e = \frac{V_{\rm bowshock} \int_{E_{\rm min}}^{E_{\rm max}} Q(E) E dE}{L_{\rm wind}} \text{ .}
\end{equation}

\noindent We assume that, as discussed above, the energy losses are dominated by escape at all electron energies. In that scenario, as escape losses are energy independent, energy should be injected into the electron population at a rate $Q(E) = N(E)/t_{\rm esc}$. As we saw in the first section of this appendix, we can link the normalisation $\kappa$ of the electron number density distribution to $J_{\nu}$ (Equation \ref{eq:jnu}) and therefore to the observed radio flux $S_\nu$ (via $4\pi D^2 S_{\nu} / V_{\rm bowshock}$). Finally, the available kinetic wind power is given by the volume filling factor of the bow shock times the total kinetic power:

\begin{equation}
    L_{\rm wind} = \frac{\eta_{\rm vol} \dot{M}_{\rm wind}  v^2_{\infty}}{2} \equiv \frac{3V_{\rm bowshock}}{4\pi R_0^3}\frac{\dot{M}_{\rm wind} v^2_{\infty}}{2} \text{ .}
\end{equation}

\noindent Combining all equations above then yields Equation \ref{eq:eta} for the injection efficiency. As stated in Footnote 8 of the main text, we simplify the relation between $Q(E)$ and $N(E)$ for analytic purposes. However, one can qualitatively imagine how including additional cooling mechanisms will increase the rate at which energy should be injected, in order to maintain a steady-state. As the wind power remains constant, including other cooling processes therefore implies a higher injection efficiency. 


\bsp	
\label{lastpage}

\begin{thebibliography}{}
\makeatletter
\relax
\def\mn@urlcharsother{\let\do\@makeother \do\$\do\&\do\#\do\^\do\_\do\%\do\~}
\def\mn@doi{\begingroup\mn@urlcharsother \@ifnextchar [ {\mn@doi@}
  {\mn@doi@[]}}
\def\mn@doi@[#1]#2{\def\@tempa{#1}\ifx\@tempa\@empty \href
  {http://dx.doi.org/#2} {doi:#2}\else \href {http://dx.doi.org/#2} {#1}\fi
  \endgroup}
\def\mn@eprint#1#2{\mn@eprint@#1:#2::\@nil}
\def\mn@eprint@arXiv#1{\href {http://arxiv.org/abs/#1} {{\tt arXiv:#1}}}
\def\mn@eprint@dblp#1{\href {http://dblp.uni-trier.de/rec/bibtex/#1.xml}
  {dblp:#1}}
\def\mn@eprint@#1:#2:#3:#4\@nil{\def\@tempa {#1}\def\@tempb {#2}\def\@tempc
  {#3}\ifx \@tempc \@empty \let \@tempc \@tempb \let \@tempb \@tempa \fi \ifx
  \@tempb \@empty \def\@tempb {arXiv}\fi \@ifundefined
  {mn@eprint@\@tempb}{\@tempb:\@tempc}{\expandafter \expandafter \csname
  mn@eprint@\@tempb\endcsname \expandafter{\@tempc}}}

\bibitem[\protect\citeauthoryear{{Angel}, {McGraw}  \& {Stockman}}{{Angel}
  et~al.}{1973}]{angel1973}
{Angel} J.~R.~P.,  {McGraw} J.~T.,   {Stockman} H.~S. J.,  1973, \mn@doi
  [\apjl] {10.1086/181293}, \href
  {https://ui.adsabs.harvard.edu/abs/1973ApJ...184L..79A} {184, L79}

\bibitem[\protect\citeauthoryear{{Astropy Collaboration} et~al.,}{{Astropy
  Collaboration} et~al.}{2013}]{astropy13}
{Astropy Collaboration} et~al., 2013, \mn@doi [\aap]
  {10.1051/0004-6361/201322068}, \href
  {https://ui.adsabs.harvard.edu/abs/2013A&A...558A..33A} {558, A33}

\bibitem[\protect\citeauthoryear{{Astropy Collaboration} et~al.,}{{Astropy
  Collaboration} et~al.}{2018}]{astropy18}
{Astropy Collaboration} et~al., 2018, \mn@doi [\aj] {10.3847/1538-3881/aabc4f},
  \href {https://ui.adsabs.harvard.edu/abs/2018AJ....156..123A} {156, 123}

\bibitem[\protect\citeauthoryear{{Bell}}{{Bell}}{1978a}]{bell1978a}
{Bell} A.~R.,  1978a, \mn@doi [\mnras] {10.1093/mnras/182.2.147}, \href
  {https://ui.adsabs.harvard.edu/abs/1978MNRAS.182..147B} {182, 147}

\bibitem[\protect\citeauthoryear{{Bell}}{{Bell}}{1978b}]{bell1978b}
{Bell} A.~R.,  1978b, \mn@doi [\mnras] {10.1093/mnras/182.3.443}, \href
  {https://ui.adsabs.harvard.edu/abs/1978MNRAS.182..443B} {182, 443}

\bibitem[\protect\citeauthoryear{{Benaglia}, {Romero}, {Mart{\'\i}}, {Peri}  \&
  {Araudo}}{{Benaglia} et~al.}{2010}]{benaglia2010}
{Benaglia} P.,  {Romero} G.~E.,  {Mart{\'\i}} J.,  {Peri} C.~S.,   {Araudo}
  A.~T.,  2010, \mn@doi [\aap] {10.1051/0004-6361/201015232}, \href
  {https://ui.adsabs.harvard.edu/abs/2010A&A...517L..10B} {517, L10}

\bibitem[\protect\citeauthoryear{{Benaglia}, {del Palacio}, {Hales}  \&
  {Colazo}}{{Benaglia} et~al.}{2021}]{benaglia2021}
{Benaglia} P.,  {del Palacio} S.,  {Hales} C.,   {Colazo} M.~E.,  2021, \mn@doi
  [\mnras] {10.1093/mnras/stab662}, \href
  {https://ui.adsabs.harvard.edu/abs/2021MNRAS.503.2514B} {503, 2514}

\bibitem[\protect\citeauthoryear{{Brookes}}{{Brookes}}{2016}]{brookes2016}
{Brookes} D.~P.,  2016, PhD thesis, University of Birmingham

\bibitem[\protect\citeauthoryear{{Brown} \& {Bomans}}{{Brown} \&
  {Bomans}}{2005}]{brown2005}
{Brown} D.,  {Bomans} D.~J.,  2005, \mn@doi [\aap]
  {10.1051/0004-6361:20041054}, \href
  {https://ui.adsabs.harvard.edu/abs/2005A&A...439..183B} {439, 183}

\bibitem[\protect\citeauthoryear{{Bychkov}, {Bychkova}  \& {Madej}}{{Bychkov}
  et~al.}{2009}]{bychkov2009}
{Bychkov} V.~D.,  {Bychkova} L.~V.,   {Madej} J.,  2009, \mn@doi [\mnras]
  {10.1111/j.1365-2966.2008.14227.x}, \href
  {https://ui.adsabs.harvard.edu/abs/2009MNRAS.394.1338B} {394, 1338}

\bibitem[\protect\citeauthoryear{{Castro Segura} et~al.,}{{Castro Segura}
  et~al.}{2021}]{castro2021}
{Castro Segura} N.,  et~al., 2021, \mn@doi [\mnras] {10.1093/mnras/staa2516},
  \href {https://ui.adsabs.harvard.edu/abs/2021MNRAS.501.1951C} {501, 1951}

\bibitem[\protect\citeauthoryear{{Comeron} \& {Kaper}}{{Comeron} \&
  {Kaper}}{1998}]{comeron1998}
{Comeron} F.,  {Kaper} L.,  1998, \aap, \href
  {https://ui.adsabs.harvard.edu/abs/1998A&A...338..273C} {338, 273}

\bibitem[\protect\citeauthoryear{{Condon}, {Cotton}, {Greisen}, {Yin},
  {Perley}, {Taylor}  \& {Broderick}}{{Condon} et~al.}{1998}]{condon1998}
{Condon} J.~J.,  {Cotton} W.~D.,  {Greisen} E.~W.,  {Yin} Q.~F.,  {Perley}
  R.~A.,  {Taylor} G.~B.,   {Broderick} J.~J.,  1998, \mn@doi [\aj]
  {10.1086/300337}, \href
  {https://ui.adsabs.harvard.edu/abs/1998AJ....115.1693C} {115, 1693}

\bibitem[\protect\citeauthoryear{{Cordes}, {Romani}  \& {Lundgren}}{{Cordes}
  et~al.}{1993}]{cordes1993}
{Cordes} J.~M.,  {Romani} R.~W.,   {Lundgren} S.~C.,  1993, \mn@doi [\nat]
  {10.1038/362133a0}, \href
  {https://ui.adsabs.harvard.edu/abs/1993Natur.362..133C} {362, 133}

\bibitem[\protect\citeauthoryear{{De Becker}, {del Valle}, {Romero}, {Peri}  \&
  {Benaglia}}{{De Becker} et~al.}{2017}]{debecker2017}
{De Becker} M.,  {del Valle} M.~V.,  {Romero} G.~E.,  {Peri} C.~S.,
  {Benaglia} P.,  2017, \mn@doi [\mnras] {10.1093/mnras/stx1826}, \href
  {https://ui.adsabs.harvard.edu/abs/2017MNRAS.471.4452D} {471, 4452}

\bibitem[\protect\citeauthoryear{{Draine}}{{Draine}}{1981}]{draine1981}
{Draine} B.~T.,  1981, \mn@doi [\apj] {10.1086/158864}, \href
  {https://ui.adsabs.harvard.edu/abs/1981ApJ...245..880D} {245, 880}

\bibitem[\protect\citeauthoryear{{Drury}}{{Drury}}{1983}]{drury1983}
{Drury} L.,  1983, \mn@doi [\ssr] {10.1007/BF00171901}, \href
  {https://ui.adsabs.harvard.edu/abs/1983SSRv...36...57D} {36, 57}

\bibitem[\protect\citeauthoryear{{Falanga}, {Bozzo}, {Lutovinov},
  {Bonnet-Bidaud}, {Fetisova}  \& {Puls}}{{Falanga} et~al.}{2015}]{falanga2015}
{Falanga} M.,  {Bozzo} E.,  {Lutovinov} A.,  {Bonnet-Bidaud} J.~M.,  {Fetisova}
  Y.,   {Puls} J.,  2015, \mn@doi [\aap] {10.1051/0004-6361/201425191}, \href
  {https://ui.adsabs.harvard.edu/abs/2015A&A...577A.130F} {577, A130}

\bibitem[\protect\citeauthoryear{{Fender} et~al.,}{{Fender}
  et~al.}{2016}]{fender2017}
{Fender} R.,  et~al., 2016, in MeerKAT Science: On the Pathway to the SKA.
  p.~13 (\mn@eprint {arXiv} {1711.04132})

\bibitem[\protect\citeauthoryear{{Fruscione} et~al.,}{{Fruscione}
  et~al.}{2006}]{fruscione2006}
{Fruscione} A.,  et~al., 2006, in {Silva} D.~R.,  {Doxsey} R.~E.,  eds,
  Society of Photo-Optical Instrumentation Engineers (SPIE) Conference Series
  Vol. 6270, Society of Photo-Optical Instrumentation Engineers (SPIE)
  Conference Series. p. 62701V, \mn@doi{10.1117/12.671760}

\bibitem[\protect\citeauthoryear{{Gaensler}, {Stappers}, {Frail}, {Moffett},
  {Johnston}  \& {Chatterjee}}{{Gaensler} et~al.}{2000}]{gaensler2000}
{Gaensler} B.~M.,  {Stappers} B.~W.,  {Frail} D.~A.,  {Moffett} D.~A.,
  {Johnston} S.,   {Chatterjee} S.,  2000, \mn@doi [\mnras]
  {10.1046/j.1365-8711.2000.03626.x}, \href
  {https://ui.adsabs.harvard.edu/abs/2000MNRAS.318...58G} {318, 58}

\bibitem[\protect\citeauthoryear{{Grinberg} et~al.,}{{Grinberg}
  et~al.}{2017}]{grinberg2017}
{Grinberg} V.,  et~al., 2017, \mn@doi [\aap] {10.1051/0004-6361/201731843},
  \href {https://ui.adsabs.harvard.edu/abs/2017A&A...608A.143G} {608, A143}

\bibitem[\protect\citeauthoryear{{Gull} \& {Sofia}}{{Gull} \&
  {Sofia}}{1979}]{gull1979}
{Gull} T.~R.,  {Sofia} S.,  1979, \mn@doi [\apj] {10.1086/157137}, \href
  {https://ui.adsabs.harvard.edu/abs/1979ApJ...230..782G} {230, 782}

\bibitem[\protect\citeauthoryear{{Gvaramadze}, {R{\"o}ser}, {Scholz}  \&
  {Schilbach}}{{Gvaramadze} et~al.}{2011}]{gvaramadze2011}
{Gvaramadze} V.~V.,  {R{\"o}ser} S.,  {Scholz} R.~D.,   {Schilbach} E.,  2011,
  \mn@doi [\aap] {10.1051/0004-6361/201016256}, \href
  {https://ui.adsabs.harvard.edu/abs/2011A&A...529A..14G} {529, A14}

\bibitem[\protect\citeauthoryear{{Gvaramadze}, {Alexashov}, {Katushkina}  \&
  {Kniazev}}{{Gvaramadze} et~al.}{2018}]{gvaramadze2018}
{Gvaramadze} V.~V.,  {Alexashov} D.~B.,  {Katushkina} O.~A.,   {Kniazev} A.~Y.,
   2018, \mn@doi [\mnras] {10.1093/mnras/stx3089}, \href
  {https://ui.adsabs.harvard.edu/abs/2018MNRAS.474.4421G} {474, 4421}

\bibitem[\protect\citeauthoryear{{H.~E.~S.~S. Collaboration}
  et~al.,}{{H.~E.~S.~S. Collaboration} et~al.}{2018}]{hess2018}
{H.~E.~S.~S. Collaboration} et~al., 2018, \mn@doi [\aap]
  {10.1051/0004-6361/201630151}, \href
  {https://ui.adsabs.harvard.edu/abs/2018A&A...612A..12H} {612, A12}

\bibitem[\protect\citeauthoryear{{Helder} et~al.,}{{Helder}
  et~al.}{2009}]{helder2009}
{Helder} E.~A.,  et~al., 2009, \mn@doi [Science] {10.1126/science.1173383},
  \href {https://ui.adsabs.harvard.edu/abs/2009Sci...325..719H} {325, 719}

\bibitem[\protect\citeauthoryear{{Heywood}}{{Heywood}}{2020}]{heywood2020}
{Heywood} I.,  2020, {oxkat: Semi-automated imaging of MeerKAT observations}
  (\mn@eprint {ascl} {2009.003})

\bibitem[\protect\citeauthoryear{Hunter}{Hunter}{2007}]{hunter07}
Hunter J.~D.,  2007, \mn@doi [Computing in Science \& Engineering]
  {10.1109/MCSE.2007.55}, 9, 90

\bibitem[\protect\citeauthoryear{{Joye} \& {Mandel}}{{Joye} \&
  {Mandel}}{2003}]{joye2003}
{Joye} W.~A.,  {Mandel} E.,  2003, in {Payne} H.~E.,  {Jedrzejewski} R.~I.,
  {Hook} R.~N.,  eds,  Astronomical Society of the Pacific Conference Series
  Vol. 295, Astronomical Data Analysis Software and Systems XII. p.~489

\bibitem[\protect\citeauthoryear{{Kaper}, {van Loon}, {Augusteijn},
  {Goudfrooij}, {Patat}, {Waters}  \& {Zijlstra}}{{Kaper}
  et~al.}{1997}]{kaper1997}
{Kaper} L.,  {van Loon} J.~T.,  {Augusteijn} T.,  {Goudfrooij} P.,  {Patat} F.,
   {Waters} L.~B.~F.~M.,   {Zijlstra} A.~A.,  1997, \mn@doi [\apjl]
  {10.1086/310454}, \href
  {https://ui.adsabs.harvard.edu/abs/1997ApJ...475L..37K} {475, L37}

\bibitem[\protect\citeauthoryear{{Kemp} \& {Wolstencroft}}{{Kemp} \&
  {Wolstencroft}}{1973}]{kemp1973}
{Kemp} J.~C.,  {Wolstencroft} R.~D.,  1973, \mn@doi [\apjl] {10.1086/181313},
  \href {https://ui.adsabs.harvard.edu/abs/1973ApJ...185L..21K} {185, L21}

\bibitem[\protect\citeauthoryear{{Kenyon}, {Smirnov}, {Grobler}  \&
  {Perkins}}{{Kenyon} et~al.}{2018}]{kenyon2018}
{Kenyon} J.~S.,  {Smirnov} O.~M.,  {Grobler} T.~L.,   {Perkins} S.~J.,  2018,
  \mn@doi [\mnras] {10.1093/mnras/sty1221}, \href
  {https://ui.adsabs.harvard.edu/abs/2018MNRAS.478.2399K} {478, 2399}

\bibitem[\protect\citeauthoryear{{Khangulyan}, {Aharonian}  \&
  {Kelner}}{{Khangulyan} et~al.}{2014}]{khangulyan2014}
{Khangulyan} D.,  {Aharonian} F.~A.,   {Kelner} S.~R.,  2014, \mn@doi [\apj]
  {10.1088/0004-637X/783/2/100}, \href
  {https://ui.adsabs.harvard.edu/abs/2014ApJ...783..100K} {783, 100}

\bibitem[\protect\citeauthoryear{{Kobulnicky}, {Gilbert}  \&
  {Kiminki}}{{Kobulnicky} et~al.}{2010}]{kobulnicky2010}
{Kobulnicky} H.~A.,  {Gilbert} I.~J.,   {Kiminki} D.~C.,  2010, \mn@doi [\apj]
  {10.1088/0004-637X/710/1/549}, \href
  {https://ui.adsabs.harvard.edu/abs/2010ApJ...710..549K} {710, 549}

\bibitem[\protect\citeauthoryear{{Kobulnicky} et~al.,}{{Kobulnicky}
  et~al.}{2016}]{kobulnicky2016}
{Kobulnicky} H.~A.,  et~al., 2016, \mn@doi [\apjs]
  {10.3847/0067-0049/227/2/18}, \href
  {https://ui.adsabs.harvard.edu/abs/2016ApJS..227...18K} {227, 18}

\bibitem[\protect\citeauthoryear{{Kretschmar} et~al.,}{{Kretschmar}
  et~al.}{2021}]{kretschmar2021}
{Kretschmar} P.,  et~al., 2021, \mn@doi [\aap] {10.1051/0004-6361/202040272},
  \href {https://ui.adsabs.harvard.edu/abs/2021A&A...652A..95K} {652, A95}

\bibitem[\protect\citeauthoryear{{Landau} \& {Lifshitz}}{{Landau} \&
  {Lifshitz}}{1959}]{landau1959}
{Landau} L.~D.,  {Lifshitz} E.~M.,  1959, {Fluid mechanics}

\bibitem[\protect\citeauthoryear{{Lequeux}}{{Lequeux}}{2005}]{lequeux2005}
{Lequeux} J.,  2005, {The Interstellar Medium}, \mn@doi{10.1007/b137959.
}

\bibitem[\protect\citeauthoryear{{Longair}}{{Longair}}{2011}]{longair2011}
{Longair} M.~S.,  2011, {High Energy Astrophysics}

\bibitem[\protect\citeauthoryear{{L{\'o}pez-Santiago}
  et~al.,}{{L{\'o}pez-Santiago} et~al.}{2012}]{lopezsantiago2012}
{L{\'o}pez-Santiago} J.,  et~al., 2012, \mn@doi [\apjl]
  {10.1088/2041-8205/757/1/L6}, \href
  {https://ui.adsabs.harvard.edu/abs/2012ApJ...757L...6L} {757, L6}

\bibitem[\protect\citeauthoryear{{Ma{\'\i}z Apell{\'a}niz}, {Pantaleoni
  Gonz{\'a}lez}, {Barb{\'a}}, {Sim{\'o}n-D{\'\i}az}, {Negueruela}, {Lennon},
  {Sota}  \& {Trigueros P{\'a}ez}}{{Ma{\'\i}z Apell{\'a}niz}
  et~al.}{2018}]{maizapellaniz2018}
{Ma{\'\i}z Apell{\'a}niz} J.,  {Pantaleoni Gonz{\'a}lez} M.,  {Barb{\'a}}
  R.~H.,  {Sim{\'o}n-D{\'\i}az} S.,  {Negueruela} I.,  {Lennon} D.~J.,  {Sota}
  A.,   {Trigueros P{\'a}ez} E.,  2018, \mn@doi [\aap]
  {10.1051/0004-6361/201832787}, \href
  {https://ui.adsabs.harvard.edu/abs/2018A&A...616A.149M} {616, A149}

\bibitem[\protect\citeauthoryear{{Matthews}, {Bell}  \& {Blundell}}{{Matthews}
  et~al.}{2020}]{matthews2020}
{Matthews} J.~H.,  {Bell} A.~R.,   {Blundell} K.~M.,  2020, \mn@doi [\nar]
  {10.1016/j.newar.2020.101543}, \href
  {https://ui.adsabs.harvard.edu/abs/2020NewAR..8901543M} {89, 101543}

\bibitem[\protect\citeauthoryear{{McConnell} et~al.,}{{McConnell}
  et~al.}{2020}]{mcconnell2020}
{McConnell} D.,  et~al., 2020, \mn@doi [\pasa] {10.1017/pasa.2020.41}, \href
  {https://ui.adsabs.harvard.edu/abs/2020PASA...37...48M} {37, e048}

\bibitem[\protect\citeauthoryear{{McMullin}, {Waters}, {Schiebel}, {Young}  \&
  {Golap}}{{McMullin} et~al.}{2007}]{mcmullin2007}
{McMullin} J.~P.,  {Waters} B.,  {Schiebel} D.,  {Young} W.,   {Golap} K.,
  2007, in {Shaw} R.~A.,  {Hill} F.,   {Bell} D.~J.,  eds,  Astronomical
  Society of the Pacific Conference Series Vol. 376, Astronomical Data Analysis
  Software and Systems XVI. p.~127

\bibitem[\protect\citeauthoryear{{Meyer}, {van Marle}, {Kuiper}  \&
  {Kley}}{{Meyer} et~al.}{2016}]{meyer2016}
{Meyer} D.~M.~A.,  {van Marle} A.~J.,  {Kuiper} R.,   {Kley} W.,  2016, \mn@doi
  [\mnras] {10.1093/mnras/stw651}, \href
  {https://ui.adsabs.harvard.edu/abs/2016MNRAS.459.1146M} {459, 1146}

\bibitem[\protect\citeauthoryear{{Mohamed}, {Mackey}  \& {Langer}}{{Mohamed}
  et~al.}{2012}]{mohamed2012}
{Mohamed} S.,  {Mackey} J.,   {Langer} N.,  2012, \mn@doi [\aap]
  {10.1051/0004-6361/201118002}, \href
  {https://ui.adsabs.harvard.edu/abs/2012A&A...541A...1M} {541, A1}

\bibitem[\protect\citeauthoryear{{Noriega-Crespo}, {van Buren}  \&
  {Dgani}}{{Noriega-Crespo} et~al.}{1997}]{noriega1997}
{Noriega-Crespo} A.,  {van Buren} D.,   {Dgani} R.,  1997, \mn@doi [\aj]
  {10.1086/118298}, \href
  {https://ui.adsabs.harvard.edu/abs/1997AJ....113..780N} {113, 780}

\bibitem[\protect\citeauthoryear{{Offringa} et~al.,}{{Offringa}
  et~al.}{2014}]{offringa2014}
{Offringa} A.~R.,  et~al., 2014, \mn@doi [\mnras] {10.1093/mnras/stu1368},
  \href {https://ui.adsabs.harvard.edu/abs/2014MNRAS.444..606O} {444, 606}

\bibitem[\protect\citeauthoryear{{Oliphant}}{{Oliphant}}{2006}]{oliphant_numpy}
{Oliphant} T.~E.,  2006, {A guide to NumPy}.
Trelgol Publishing, p.~85

\bibitem[\protect\citeauthoryear{{Peri}, {Benaglia}, {Brookes}, {Stevens}  \&
  {Isequilla}}{{Peri} et~al.}{2012}]{peri2012}
{Peri} C.~S.,  {Benaglia} P.,  {Brookes} D.~P.,  {Stevens} I.~R.,   {Isequilla}
  N.~L.,  2012, \mn@doi [\aap] {10.1051/0004-6361/201118116}, \href
  {https://ui.adsabs.harvard.edu/abs/2012A&A...538A.108P} {538, A108}

\bibitem[\protect\citeauthoryear{{Peri}, {Benaglia}  \& {Isequilla}}{{Peri}
  et~al.}{2015}]{peri2015}
{Peri} C.~S.,  {Benaglia} P.,   {Isequilla} N.~L.,  2015, \mn@doi [\aap]
  {10.1051/0004-6361/201424676}, \href
  {https://ui.adsabs.harvard.edu/abs/2015A&A...578A..45P} {578, A45}

\bibitem[\protect\citeauthoryear{{Rangelov}, {Montmerle}, {Federman},
  {Boiss{\'e}}  \& {Gabici}}{{Rangelov} et~al.}{2019}]{rangelov2019}
{Rangelov} B.,  {Montmerle} T.,  {Federman} S.~R.,  {Boiss{\'e}} P.,   {Gabici}
  S.,  2019, \mn@doi [\apj] {10.3847/1538-4357/ab43e5}, \href
  {https://ui.adsabs.harvard.edu/abs/2019ApJ...885..105R} {885, 105}

\bibitem[\protect\citeauthoryear{{Robitaille} \& {Bressert}}{{Robitaille} \&
  {Bressert}}{2012}]{robitaille12}
{Robitaille} T.,  {Bressert} E.,  2012, {APLpy: Astronomical Plotting Library
  in Python} (\mn@eprint {ascl} {1208.017})

\bibitem[\protect\citeauthoryear{{Russell}, {Fender}, {Gallo}  \&
  {Kaiser}}{{Russell} et~al.}{2007}]{russell2007}
{Russell} D.~M.,  {Fender} R.~P.,  {Gallo} E.,   {Kaiser} C.~R.,  2007, \mn@doi
  [\mnras] {10.1111/j.1365-2966.2007.11539.x}, \href
  {https://ui.adsabs.harvard.edu/abs/2007MNRAS.376.1341R} {376, 1341}

\bibitem[\protect\citeauthoryear{{S{\'a}nchez-Ayaso}, {del Valle},
  {Mart{\'\i}}, {Romero}  \& {Luque-Escamilla}}{{S{\'a}nchez-Ayaso}
  et~al.}{2018}]{sanchezayaso2018}
{S{\'a}nchez-Ayaso} E.,  {del Valle} M.~V.,  {Mart{\'\i}} J.,  {Romero} G.~E.,
   {Luque-Escamilla} P.~L.,  2018, \mn@doi [\apj] {10.3847/1538-4357/aac7c7},
  \href {https://ui.adsabs.harvard.edu/abs/2018ApJ...861...32S} {861, 32}

\bibitem[\protect\citeauthoryear{{Schulz}, {Ackermann}, {Buehler}, {Mayer}  \&
  {Klepser}}{{Schulz} et~al.}{2014}]{schulz2014}
{Schulz} A.,  {Ackermann} M.,  {Buehler} R.,  {Mayer} M.,   {Klepser} S.,
  2014, \mn@doi [\aap] {10.1051/0004-6361/201423468}, \href
  {https://ui.adsabs.harvard.edu/abs/2014A&A...565A..95S} {565, A95}

\bibitem[\protect\citeauthoryear{{Smirnov} \& {Tasse}}{{Smirnov} \&
  {Tasse}}{2015}]{smirnov2018}
{Smirnov} O.~M.,  {Tasse} C.,  2015, \mn@doi [\mnras] {10.1093/mnras/stv418},
  \href {https://ui.adsabs.harvard.edu/abs/2015MNRAS.449.2668S} {449, 2668}

\bibitem[\protect\citeauthoryear{{Stappers}, {Gaensler}, {Kaspi}, {van der
  Klis}  \& {Lewin}}{{Stappers} et~al.}{2003}]{stappers2003}
{Stappers} B.~W.,  {Gaensler} B.~M.,  {Kaspi} V.~M.,  {van der Klis} M.,
  {Lewin} W.~H.~G.,  2003, \mn@doi [Science] {10.1126/science.1079841}, \href
  {https://ui.adsabs.harvard.edu/abs/2003Sci...299.1372S} {299, 1372}

\bibitem[\protect\citeauthoryear{{Tasse} et~al.,}{{Tasse}
  et~al.}{2018}]{tasse2018}
{Tasse} C.,  et~al., 2018, \mn@doi [\aap] {10.1051/0004-6361/201731474}, \href
  {https://ui.adsabs.harvard.edu/abs/2018A&A...611A..87T} {611, A87}

\bibitem[\protect\citeauthoryear{{Terada}, {Tashiro}, {Bamba}, {Yamazaki},
  {Kouzu}, {Koyama}  \& {Seta}}{{Terada} et~al.}{2012}]{terada2012}
{Terada} Y.,  {Tashiro} M.~S.,  {Bamba} A.,  {Yamazaki} R.,  {Kouzu} T.,
  {Koyama} S.,   {Seta} H.,  2012, \mn@doi [\pasj] {10.1093/pasj/64.6.138},
  \href {https://ui.adsabs.harvard.edu/abs/2012PASJ...64..138T} {64, 138}

\bibitem[\protect\citeauthoryear{{Toal{\'a}}, {Oskinova},
  {Gonz{\'a}lez-Gal{\'a}n}, {Guerrero}, {Ignace}  \& {Pohl}}{{Toal{\'a}}
  et~al.}{2016}]{toala2016}
{Toal{\'a}} J.~A.,  {Oskinova} L.~M.,  {Gonz{\'a}lez-Gal{\'a}n} A.,  {Guerrero}
  M.~A.,  {Ignace} R.,   {Pohl} M.,  2016, \mn@doi [\apj]
  {10.3847/0004-637X/821/2/79}, \href
  {https://ui.adsabs.harvard.edu/abs/2016ApJ...821...79T} {821, 79}

\bibitem[\protect\citeauthoryear{{Toal{\'a}}, {Oskinova}  \&
  {Ignace}}{{Toal{\'a}} et~al.}{2017}]{toala2017}
{Toal{\'a}} J.~A.,  {Oskinova} L.~M.,   {Ignace} R.,  2017, \mn@doi [\apjl]
  {10.3847/2041-8213/aa667c}, \href
  {https://ui.adsabs.harvard.edu/abs/2017ApJ...838L..19T} {838, L19}

\bibitem[\protect\citeauthoryear{{Watanabe} et~al.,}{{Watanabe}
  et~al.}{2006}]{watanabe2006}
{Watanabe} S.,  et~al., 2006, \mn@doi [\apj] {10.1086/507458}, \href
  {https://ui.adsabs.harvard.edu/abs/2006ApJ...651..421W} {651, 421}

\bibitem[\protect\citeauthoryear{{Wiersema} et~al.,}{{Wiersema}
  et~al.}{2009}]{wiersema2009}
{Wiersema} K.,  et~al., 2009, \mn@doi [\mnras]
  {10.1111/j.1745-3933.2009.00643.x}, \href
  {https://ui.adsabs.harvard.edu/abs/2009MNRAS.397L...6W} {397, L6}

\bibitem[\protect\citeauthoryear{{Wilkin}}{{Wilkin}}{1996}]{wilkin1996}
{Wilkin} F.~P.,  1996, \mn@doi [\apjl] {10.1086/309939}, \href
  {https://ui.adsabs.harvard.edu/abs/1996ApJ...459L..31W} {459, L31}

\bibitem[\protect\citeauthoryear{{del Palacio}, {Bosch-Ramon}, {M{\"u}ller}  \&
  {Romero}}{{del Palacio} et~al.}{2018}]{delpalacio2018}
{del Palacio} S.,  {Bosch-Ramon} V.,  {M{\"u}ller} A.~L.,   {Romero} G.~E.,
  2018, \mn@doi [\aap] {10.1051/0004-6361/201833321}, \href
  {https://ui.adsabs.harvard.edu/abs/2018A&A...617A..13D} {617, A13}

\bibitem[\protect\citeauthoryear{{del Valle} \& {Pohl}}{{del Valle} \&
  {Pohl}}{2018}]{delvalle2018}
{del Valle} M.~V.,  {Pohl} M.,  2018, \mn@doi [\apj]
  {10.3847/1538-4357/aad333}, \href
  {https://ui.adsabs.harvard.edu/abs/2018ApJ...864...19D} {864, 19}

\bibitem[\protect\citeauthoryear{{del Valle} \& {Romero}}{{del Valle} \&
  {Romero}}{2012}]{delvalle2012}
{del Valle} M.~V.,  {Romero} G.~E.,  2012, \mn@doi [\aap]
  {10.1051/0004-6361/201218937}, \href
  {https://ui.adsabs.harvard.edu/abs/2012A&A...543A..56D} {543, A56}

\bibitem[\protect\citeauthoryear{{del Valle}, {Romero}  \& {De Becker}}{{del
  Valle} et~al.}{2013}]{delvalle2013}
{del Valle} M.~V.,  {Romero} G.~E.,   {De Becker} M.,  2013, \mn@doi [\aap]
  {10.1051/0004-6361/201220112}, \href
  {https://ui.adsabs.harvard.edu/abs/2013A&A...550A.112D} {550, A112}

\makeatother
\end{thebibliography}
\end{document}